\documentclass[preprintnumbers,prd,superscriptaddress,nofootinbib]{revtex4}

\usepackage{amsfonts}
\usepackage{amssymb}
\usepackage{graphicx}

\usepackage[titletoc]{appendix}
\usepackage{color}
\usepackage[rightcaption]{sidecap}
\usepackage{subfigure}

\usepackage{amsmath}
\usepackage{dcolumn}
\usepackage{array}
\usepackage{ctable}
\usepackage{multirow}
\usepackage{siunitx}
\usepackage{longtable}
\usepackage{tabularx}
\usepackage{booktabs}

\def\be{\begin{equation}}
\def\ee{\end{equation}}
\def\bea{\begin{eqnarray}}
\def\eea{\end{eqnarray}}

\begin{document}

\title{Testing dark matter density profiles based on Pad\'e approximants of different orders}

\author{Talgar~\surname{Konysbayev}}
\email[] {talgar_777@mail.ru}
\affiliation{National Nanotechnology Laboratory of Open Type,  Almaty 050040, Kazakhstan.}
\affiliation{Farabi University, Al-Farabi av. 71, 050040 Almaty, Kazakhstan.}

\author{Orlando Luongo}
\email[]{orlando.luongo@unicam.it}
\affiliation{Farabi University, Al-Farabi av. 71, 050040 Almaty, Kazakhstan.}
\affiliation{Dipartimento di Matematica, Universit\`a di Pisa, Largo B. Pontecorvo, 56127 Pisa, Italy.}
\affiliation{Universit\`a di Camerino, Divisione di Fisica, Via Madonna delle carceri, 62032 Camerino, Italy.}

\author{Marco~\surname{Muccino}}
\email[]{marco.muccino@unicam.it}
\affiliation{Farabi University, Al-Farabi av. 71, 050040 Almaty, Kazakhstan.}
\affiliation{Universit\`a di Camerino, Divisione di Fisica, Via Madonna delle carceri, 62032 Camerino, Italy.}
\affiliation{ICRANet, Piazza della Repubblica 10, 65122 Pescara, Italy.}

\author{Yergali~\surname{Kurmanov}}
\email[]{kurmanov.yergali@kaznu.kz}
\affiliation{National Nanotechnology Laboratory of Open Type,  Almaty 050040, Kazakhstan.}
\affiliation{Farabi University, Al-Farabi av. 71, 050040 Almaty, Kazakhstan.}

\author{Kuantay~\surname{Boshkayev}}
\email[]{kuantay@mail.ru}
\affiliation{National Nanotechnology Laboratory of Open Type,  Almaty 050040, Kazakhstan.}
\affiliation{Farabi University, Al-Farabi av. 71, 050040 Almaty, Kazakhstan.}

\author{ Gulnara~\surname{Suliyeva}}
\email[]{g_suliyeva@mail.ru}
\affiliation{National Nanotechnology Laboratory of Open Type,  Almaty 050040, Kazakhstan.}
\affiliation{Farabi University, Al-Farabi av. 71, 050040 Almaty, Kazakhstan.}
\affiliation{Fesenkov Astrophysical Institute, Observatory 23, 050020 Almaty, Kazakhstan.}

\author{Ainur~\surname{Urazalina}}
\email[]{y.a.a.707@mail.ru}
\affiliation{National Nanotechnology Laboratory of Open Type,  Almaty 050040, Kazakhstan.}
\affiliation{Farabi University, Al-Farabi av. 71, 050040 Almaty, Kazakhstan.}

\author{Guldana~\surname{Rabigulova}}
\email[]{guldanaberikhanovna@gmail.com}
\affiliation{National Nanotechnology Laboratory of Open Type,  Almaty 050040, Kazakhstan.}
\affiliation{Farabi University, Al-Farabi av. 71, 050040 Almaty, Kazakhstan.}

\begin{abstract}
We investigate whether low-order Pad\'e rational functions can be used as empirical density profiles for modeling dark matter halos from galaxy RCs. We introduce three parameterizations determined from Pad\'e series, denoted Pad\'e 02, Pad\'e 12 and Pad\'e 03, and compare them with pseudo-isothermal, Burkert, Beta, Brownstein, exponential-sphere and Persic models. The analysis is performed for eight galaxies  under the assumption that RCs are dark matter dominated. The parameters are inferred from RC data and the relative statistical performance of the models is obtained  through the Bayesian Information Criterion. We find that the Pad\'e profiles perform well, being comparable with the other profiles, albeit not universally preferred. In most galaxies, conventional two parameter profiles provide fits of comparable or better statistical quality, whereas the clearest improvement occurs for one particular galaxy, where the Pad\'e 03 profile gives the lowest BIC. Even though the Pad\'e profiles are often statistically less strong than other models, they appear plausible in explaining the dark matter nature. Hence, their empirical construction can therefore be used to reconstruct RCs phenomenologically. 
\end{abstract}

\maketitle

\section{Introduction}

Dark matter (DM) distribution in galaxies remains an open problem in modern astrophysics, with evident consequences in cosmology. Nevertheless, evidence for a dominant non-luminous component comes from a variety of independent observations, including galaxy and cluster dynamics, gravitational lensing, the cosmic microwave background, and the formation of large-scale structure \cite{1933AcHPh...6..110Z,2017arXiv171101693A,Rubin,2006ApJ...648L.109C,Plank2013,2020A&A...641A...1P}. 

At galactic scales, one of the most direct probes of the DM distribution is provided by RCs (RCs). If the mass distribution were dominated by the luminous disk alone, the circular velocity would be expected to decrease approximately as $v(r)\propto r^{-1/2}$ outside the baryonic region. Observed RCs, however, often remain nearly flat or decline much more slowly at large radii, implying that the enclosed gravitational mass continues to increase well beyond the luminous component \cite{Rubin,1981AJ.....86.1791B,1981AJ.....86.1825B,1991MNRAS.249..523B}.

The standard interpretation is that galaxies are embedded in extended DM halos. A large number of analytic density profiles has been proposed to describe such halos. The Navarro-Frenk-White (NFW) profile, motivated by cold DM $N$-body simulations, predicts a cusped central density distribution \cite{1995MNRAS.275..720N}. The Moore profile gives an even steeper inner cusp \cite{Moore}, while the Einasto profile introduces an additional shape parameter and is widely used in numerical studies because of its flexibility\footnote{The profile is particularly interesting since it appears under the form of an exponential.} \cite{1965TrAlm...5...87E,Merritt,2010MNRAS.402...21N}. On the other hand, cored phenomenological profiles, such as the pseudo-isothermal (ISO) and Burkert profiles, are often successful in dwarf and low-surface-brightness galaxies, where the inner RCs tend to favor finite central densities over steep cusps \cite{1991MNRAS.249..523B,Burkert,Blok}. Other empirical parameterizations, including the exponential sphere, the Beta and Brownstein profiles, as well as the universal RC of Persic, Salucci and Stel, have also been used in different galactic samples and mass ranges \cite{1996MNRAS.281...27P,2006ApJ...636..721B,Sofue2013,2020Galax...8...37S}. Related analyses have been carried out for dwarf galaxies \cite{2004ApJ...617.1059R,2009A&A...493..871S,2015MNRAS.452.3650O,2001MNRAS.325.1017V,2016MNRAS.462.3628R,2017MNRAS.465.4703K,2018MNRAS.473.4392S,2025Ap&SS.370..129O}, low-surface-brightness galaxies \cite{2000AJ....119.1579V,2005ApJ...634..227D,2004MNRAS.355..794H,2019MNRAS.490.5451D,2008ApJ...676..920K}, and spiral galaxies \cite{2017PASJ...69R...1S,boshkayev2020,2020Galax...8...74B,2021MNRAS.508.1543B,boshkayev2023analysis,2024IJMPD..3350016B,2025RAA....25g5005K}.

Despite this variety of models, RC fitting is not a unique inverse problem. Indeed, \emph{distinct density laws may reproduce the same data with comparable statistical quality while leading to different values of the central density, scale radius, and halo mass}. This degeneracy is well known in disk-halo decompositions and becomes particularly relevant when the baryonic contribution is poorly constrained or when the data cover only a limited radial interval \cite{1985ApJ...295..305V,2001AJ....122.2396D,2004MNRAS.351..903G,2008AJ....136.2648D}. Uncertainties in the stellar mass-to-light ratio, gas contribution, and disk geometry can propagate directly into the inferred halo parameters \cite{1986RSPTA.320..447V,2011AJ....142...24O}. 

For the above reason, it is useful to test alternative analytic parameterizations that are flexible enough to describe different radial regimes and, at the same time, simple enough to preserve closed-form expressions characterizing spiral galaxies. 

Remarkably, rational approximants provide a natural framework for this purpose. Pad\'e approximants are commonly used in mathematical physics to represent a function by a ratio of polynomials, often improving the behavior of truncated series and extending their range of validity \cite{1981pabt.book.....B,1981paea.book.....B}. In gravitational physics and cosmology, Pad\'e-type constructions have been used to re-sum perturbative expansions, improve post-Newtonian approximations, approximate luminosity distances and build accurate analytic representations of cosmological observables \cite{1998PhRvD..57..885D,2000PhRvD..62d4024D,2014PhRvD..89j3506G,2014PhRvD..90d3531A,2014JCAP...01..045W}. 

Interestingly, an analogous strategy can be applied at the level of galactic halo profiles: \emph{a low-order rational density law may interpolate between the inner and outer regions of a halo with only a small number of parameters}, without characterizing a DM model \emph{a priori}.

Motivated by the above considerations, we here test this possibility phenomenologically by employing the Pad\'e expansions. To do so, we introduce three Pad\'e-inspired density profiles and apply them to the RCs of the following galaxies: ESO0140040, ESO3050090, ESO4250180, ESO4880049, U5750, U11454, U11748, and U11819. The underlying galaxies have been selected since they have publicly available RC data and have been used in previous studies of phenomenological DM halo profiles. Moreover, we ensure within them that the main component responsible for the RCs is the DM constituent, neglecting at first glance, the role of gas and baryons. Accordingly, such galaxies provide a useful test set spanning different velocity amplitudes and radial coverages, while remaining sufficiently small to allow a detailed comparison among several standard and Pad\'e-inspired parameterizations. The proposed profiles are thus compared with standard phenomenological halo models, namely the ISO, Burkert, Beta, Brownstein, exponential-sphere, and Persic parameterizations. Our comparison is based on the Bayesian Information Criterion (BIC), so that possible improvements in the likelihood are weighed against the larger number of free parameters in the Pad\'e models. In our treatment, we do not aim at formulating a DM microscopic model, trying rather to obtain  empirical fitting functions that may suggest how DM halos form. Hence, their use is assessed by the corresponding reproduction of the observed RCs without introducing unnecessary parameter freedom. This reinforces the fact that we intentionally neglect the explicit stellar and gas contributions and approximate the observed RCs by the spherical halo contribution alone. As consequence of our recipe, we find an overall agreement with our Pad\'e curves, albeit the standard profiles appears better performing, with one galactic exception only. We conclude that the inclusion of additional components would be essential to check the goodness of our Pad\'e-inspired profiles, although they seem to have interesting impact in framing the RCs. 

The paper is organized as follows. In Sec.~\ref{profiles}, we review the standard phenomenological halo profiles used for comparison and introduce the Pad\'e-inspired density models. In Sec.~\ref{methods}, we describe the fitting procedure, the computation of the enclosed mass and circular velocity, and the statistical criterion adopted for model comparison. In Sec.~\ref{results}, we present the results for the ESO and U galaxy samples and discuss when the Pad\'e profiles are competitive with conventional halo models. Finally, Sec.~\ref{conclusion} summarizes our conclusions and outlines possible extensions of the analysis.

\section{Dark matter profiles}\label{profiles}

In this section, we present several widely used DM density profiles, and propose new models based on the Pad\'e approximant. We provide analytical expressions for all the mentioned models with the description of their main characteristics and application.

\subsection{Traditional Dark Matter Models}

Traditional DM models include ISO, exponential sphere, Burkert, Beta, Brownstein, as well as the velocity profile proposed by Persic. These models are commonly employed to describe the distribution of DM in galactic halos.

Most of the considered profiles are characterized by two main parameters: the central (or characteristic) DM density $\rho_0$ and the characteristic halo scale radius $r_0$; only the Persic model fits directly the velocity profile thus, instead of $\rho_0$, it fits the parameters $\beta_0$ that represents the ratio of the terminal velocity to the speed of light.
Accordingly, it appears particularly useful to consider the dimensionless radial coordinate, introduced as
$x=r/r_0$.
The standard phenomenological profiles used for comparison are collected below. 
\begin{equation}
\begin{array}{ll}
\rho_{ISO}(x)=\dfrac{\rho_0}{1+x^2}, 
& \text{ISO profile \cite{Jimenez}}, \\[2.0ex]

\rho_{Exp}(x)=\rho_0 e^{-x}, 
& \text{exponential-sphere profile \cite{Sofue2013}}, \\[2.0ex]

\rho_{Bur}(x)=\dfrac{\rho_0}{(1+x)(1+x^2)}, 
& \text{Burkert profile \cite{Burkert}}, \\[2.0ex]

\rho_{Beta}(x)=\dfrac{\rho_0}{(1+x^2)^{3/2}}, 
& \text{Beta profile \cite{1995MNRAS.275..720N,2020Galax...8...37S}}, \\[2.0ex]

\rho_{Bro}(x)=\dfrac{\rho_0}{1+x^3}, 
& \text{Brownstein profile \cite{2006ApJ...636..721B,2020Galax...8...37S}},\\[2.0ex]
\beta^2(x)=\beta_0^2\dfrac{x^2}{1+x^2},
& \text{Persic profile  \cite{Barranco,1996MNRAS.281...27P}}.
\end{array}
\label{eq:standard_profiles}
\end{equation}

\textcolor{blue}{}

\subsection{Novel Pad\'e profiles}

We now introduce the Pad\'e-inspired density profiles used in the analysis. The construction is phenomenological: the aim is not to derive a halo profile from an underlying microscopic DM model, but to test whether simple rational functions can provide flexible empirical descriptions of the observed RCs.

A Pad\'e approximant represents a function as the ratio of two polynomials \cite{Baker1996},
\begin{equation}
P_{n,m}(x)=
\frac{a_0+a_1x+\cdots+a_nx^n}
{b_0+b_1x+\cdots+b_mx^m},
\label{eq:pade_general}
\end{equation}
where $n$ and $m$ are the orders of the numerator and denominator, respectively. Motivated by this form, we consider three low-order rational density laws,

\begin{subequations}
    \begin{align}
&\rho_{02}(x)=
\rho_0\frac{1}{1+b_1x+b_2x^2},\label{rho02}\\
&\rho_{12}(x)=
\rho_0\frac{1+a_1x}{1+b_1x+b_2x^2},
\label{rho12}\\
&\rho_{03}(x)=
\rho_0\frac{1}{1+b_1x+b_2x^2+b_3x^3}.
\label{rho03}        
    \end{align}
\end{subequations}
Here $\rho_0$ is the central density scale, $x=r/r_0$, and $a_1$, $b_1$, $b_2$, and $b_3$ are dimensionless coefficients to be constrained by the RC data.

The three profiles have different outer behaviors. For $x\gg1$, the Pad\'e 02 profile scales as $\rho_{02}\sim x^{-2}$ and therefore naturally gives an approximately flat circular velocity at large radii. The Pad\'e 03 profile instead behaves as $\rho_{03}\sim x^{-3}$, corresponding to a steeper outer falloff. The Pad\'e 12 profile generically behaves as $\rho_{12}\sim x^{-1}$ when $a_1\neq0$; for this reason, it should not be extrapolated far outside the fitted radial range without additional physical constraints. Throughout this work, the Pad\'e profiles are used only as empirical fitting functions over the radial interval covered by the data.

Physical regularity is imposed directly in the fitting procedure. At each step of the Markov chain, we require the density to remain positive over the observed radial interval, $
\rho(r)>0$, and we exclude parameter values for which the denominator of the Pad\'e profile vanishes in the same interval. Parameter points violating either condition are assigned zero likelihood. These requirements ensure that the fitted profiles are regular and positive over the domain actually probed by the RC data.

The fits obtained with the standard phenomenological profiles are shown in Fig.~\ref{fig:RC1}. The data points with error bars denote the observed RCs of ESO0140040, ESO3050090, ESO4250180, ESO4880049, U5750, U11454, U11748, and U11819. The solid curves correspond to the Beta, Brownstein, Burkert, exponential-sphere, pseudo-isothermal, and Persic parameterizations. The corresponding fits obtained with the Pad\'e-inspired profiles are displayed in Fig.~\ref{fig:RC2}, where the solid curves represent the Pad\'e 02, Pad\'e 12, and Pad\'e 03 models.


\section{Methods and numerical analysis}
\label{methods}

We now describe the procedure used to fit the RCs of the galaxies reported in Table \ref{tab:galaxy_groups}.

\begin{table}[t]
\centering
\caption{Qualitative grouping of the galaxies used in the analysis. The classification is intended only to organize the discussion and does not define a statistically complete sample.}
\label{tab:galaxy_groups}
\begin{tabular}{l l}
\hline
Group & Galaxies \\
\hline
ESO subsample & ESO0140040, ESO3050090, ESO4250180, ESO4880049 \\
U subsample & U5750, U11454, U11748, U11819 \\
Standard-profile dominated & ESO3050090, ESO4250180, ESO4880049, U5750, U11454, U11819 \\
Pad\'e-sensitive case & U11748 \\
High-amplitude / structured RCs & ESO0140040, U11748 \\
\hline
\end{tabular}
\end{table}

The analysis is performed approximating DM to be the most dominant species for each of the aforementioned galaxies. This halo-only approximation clearly does not consider the other baryonic components, including gas and can be therefore considered as a first attempt to check whether the phenomenological reconstructions, using the Pad\'e profiles,  introduced in Sec.~\ref{profiles}, could work with present data. More precisely, our intention is to compare our findings with the standard phenomenological halo profiles listed above.

Since the baryonic decomposition of the selected galaxies is not included in the present analysis, we approximate the observed rotational velocity by the contribution generated by the spherical DM halo,
\begin{equation}
v_{\rm tot}^2(r)\simeq v_{\rm halo}^2(r).
\label{eq:vhalo_approx}
\end{equation}
This approximation should be understood as a phenomenological halo-dominated fit, not as a full mass decomposition. In particular, stellar-disk, gas, and bulge contributions are not modeled separately.

For a spherically symmetric density profile $\rho(r)$, the enclosed mass and circular velocity are respectively, 
\begin{subequations}
\begin{align}
&M(r)=4\pi\int_0^r \tilde r^{\,2}\rho(\tilde r)\,d\tilde r,\label{eq:mass_def}\\
&v_{\rm halo}(r)=\sqrt{\frac{G M(r)}{r}} .
\label{eq:velocity_def}
\end{align}
\end{subequations}
The mass functions for the ISO, exponential-sphere, Burkert, Beta, Brownstein, and Persic parameterizations have been discussed in Refs.~\cite{2020Galax...8...74B,2021MNRAS.508.1543B,2024IJMPD..3350016B,Suliyeva_2024,2026arXiv260423353S,boshkayev2020,boshkayev2023analysis}. Below we report the mass functions for the Pad\'e profiles whenever their analytic form is compact enough to be useful.

For the Pad\'e 02 profile, substituting Eq.~\eqref{rho02} into Eq.~\eqref{eq:mass_def} gives, for $b_1^2-4b_2>0$,
\begin{equation}
\begin{split}
M_{02}(x)=\frac{2\pi \rho_{0}r_0^3}{b_{2}^{2}\sqrt{b_{1}^{2}-4b_{2}}}
\Bigg\{
&2(b_1^2-2b_2)
\left[
\operatorname{arctanh}\left(\frac{b_1}{\sqrt{b_1^2-4b_2}}\right)
-
\operatorname{arccoth}\left(\frac{\sqrt{b_1^2-4b_2}}{2b_2x+b_1}\right)
\right]
\\
&+\sqrt{b_1^2-4b_2}
\left[
2b_2x-b_1\ln(1+b_1x+b_2x^2)
\right]
\Bigg\}.
\end{split}
\label{eq:M02_pos}
\end{equation}
For the Pad\'e 12 profile one obtains
\begin{equation}
\begin{split}
M_{12}(x)=\frac{2\pi \rho_0 r_0^3}{b_2^3\sqrt{b_1^2-4b_2}}
\Bigg\{
&2A
\left[
\operatorname{arctanh}\left(\frac{b_1}{\sqrt{b_1^2-4b_2}}\right)
-
\operatorname{arccoth}\left(\frac{\sqrt{b_1^2-4b_2}}{2b_2x+b_1}\right)
\right]
\\
&+\sqrt{b_1^2-4b_2}
\left[
-b_2x(a_1b_2x-2a_1b_1+2b_2)
-B\ln(1+b_1x+b_2x^2)
\right]
\Bigg\},
\end{split}
\label{eq:M12_pos}
\end{equation}
where
\begin{equation}
A=a_1b_1^3-b_1(3a_1+b_1)b_2+2b_2^2,
\qquad
B=a_1b_1^2-(a_1+b_1)b_2 .
\label{eq:AB_def}
\end{equation}

When $b_1^2-4b_2<0$, the same mass functions are more conveniently written in terms of inverse tangent functions. For Pad\'e 02,
\begin{equation}
\begin{split}
M_{02}(x)=\frac{2\pi \rho_{0}r_0^3}{b_{2}^{2}\sqrt{4b_{2}-b_{1}^{2}}}
\Bigg\{
&2(b_1^2-2b_2)
\left[
\arctan\left(\frac{2b_2x+b_1}{\sqrt{4b_2-b_1^2}}\right)
-
\arctan\left(\frac{b_1}{\sqrt{4b_2-b_1^2}}\right)
\right]
\\
&+\sqrt{4b_2-b_1^2}
\left[
2b_2x-b_1\ln(1+b_1x+b_2x^2)
\right]
\Bigg\},
\end{split}
\label{eq:M02_neg}
\end{equation}
whereas for Pad\'e 12,
\begin{equation}
\begin{split}
M_{12}(x)=\frac{2\pi \rho_0 r_0^3}{b_2^3\sqrt{4b_2-b_1^2}}
\Bigg\{
&2A
\left[
\arctan\left(\frac{2b_2x+b_1}{\sqrt{4b_2-b_1^2}}\right)
-
\arctan\left(\frac{b_1}{\sqrt{4b_2-b_1^2}}\right)
\right]
\\
&+\sqrt{4b_2-b_1^2}
\left[
-b_2x(a_1b_2x-2a_1b_1+2b_2)
-B\ln(1+b_1x+b_2x^2)
\right]
\Bigg\}.
\end{split}
\label{eq:M12_neg}
\end{equation}
The limiting case $b_1^2-4b_2=0$ corresponds to a double root of the quadratic denominator and is obtained by taking the appropriate limit of the expressions above.

For the Pad\'e 03 profile, an analytic expression for $M_{03}(r)$ can also be written down, but it is lengthy and does not add useful physical information. In the numerical analysis we therefore compute $M_{03}(r)$ directly from Eq.~\eqref{eq:mass_def}.

\subsection{Bounds and viability of the profiles}

In order to reduce parameter degeneracies and assign a common meaning to the scale radius, we impose the condition
\begin{equation}
\rho(r_0)=\frac{\rho_0}{2}.
\label{eq:half_density_condition}
\end{equation}
For Pad\'e 02 this gives
\begin{equation}
b_2=1-b_1,
\label{eq:b2_pade02}
\end{equation}
so that the independent parameters are $(\rho_0,r_0,b_1)$. For Pad\'e 12 one obtains
\begin{equation}
b_2=1+2a_1-b_1,
\label{eq:b2_pade12}
\end{equation}
and the independent parameters are $(\rho_0,r_0,a_1,b_1)$. For Pad\'e 03 the same condition gives
\begin{equation}
b_3=1-b_1-b_2,
\label{eq:b3_pade03}
\end{equation}
leaving $(\rho_0,r_0,b_1,b_2)$ as independent parameters.

The parameters are inferred through a Markov Chain Monte Carlo (MCMC) analysis based on the Metropolis-Hastings algorithm. For a given model with parameter vector $\boldsymbol{\theta}$, the likelihood assumes Gaussian-distributed errors on the observed rotation velocities,
\begin{equation}
\ln \mathcal{L}(\boldsymbol{\theta})
=
-\frac{1}{2}
\sum_{i=1}^{N}
\left[
\frac{\left(v_i^{\rm obs}-v_{\rm th}(r_i;\boldsymbol{\theta})\right)^2}
{\sigma_{v,i}^2}
+\ln\left(2\pi\sigma_{v,i}^2\right)
\right],
\label{eq:likelihood}
\end{equation}
where $v_i^{\rm obs}$ is the observed velocity at radius $r_i$, $\sigma_{v,i}$ is its observational uncertainty, $v_{\rm th}(r_i;\boldsymbol{\theta})$ is the theoretical halo velocity predicted by the model, and $N$ is the number of data points for the galaxy under consideration.

We adopt uniform priors on the free parameters. In particular,
\begin{subequations}
\begin{align}
&\rho_0\in[0,1000]\times10^{-3}\,M_\odot\,{\rm pc}^{-3},\\
&r_0\in[0,20]\,{\rm kpc},
\label{eq:priors_rho_r}
\end{align}
\end{subequations}
and, for the Pad\'e coefficients,
\begin{subequations}
    \begin{align}
&a_1\in[-10,10],\\
&b_1\in[-10,10],\\
&b_2\in[-10,10].
\label{eq:priors_pade}
\end{align}
\end{subequations}

At each step of the chain, parameter points leading to negative densities or to poles of the Pad\'e profile within the observed radial interval are rejected. Thus, the fitted profiles remain regular and positive over the radial domain covered by the data.

For each galaxy and each model, the chains are run for $\mathcal{O}(10^5)$ steps. The values reported in Tables~\ref{tab:ESO0140040}-\ref{tab:U11819} correspond to the maximum-likelihood points, while the quoted uncertainties are obtained from the marginalized posterior distributions. The halo mass quoted for each fit is evaluated at the outermost observed radius of the corresponding RC.

To compare the statistical performance of the different profiles, we use the BIC,
\begin{equation}
{\rm BIC}=-2\ln\mathcal{L}_{\rm max}+p\ln N,
\label{eq:BIC}
\end{equation}
where $\ln\mathcal{L}_{\rm max}$ is the maximum log-likelihood, $p$ is the number of independent fitted parameters, and $N$ is the number of data points. For each galaxy, we define
\begin{equation}
\Delta{\rm BIC}_i={\rm BIC}_i-{\rm BIC}_{\rm min},
\label{eq:deltaBIC}
\end{equation}
where ${\rm BIC}_{\rm min}$ is the lowest BIC value among the models considered for that galaxy. Models with $\Delta{\rm BIC}\lesssim2$ are statistically comparable to the best model, values $2\lesssim\Delta{\rm BIC}\lesssim6$ indicate moderate evidence against the model, and $\Delta{\rm BIC}\gtrsim6$ indicates strong evidence against it. This criterion is particularly useful here because the Pad\'e profiles generally contain more free parameters than the standard two-parameter halo models.

\section{Results and Discussion}\label{results}

In this section we present our main results providing technical details and discussions.
For all the considered DM profiles, the total mass is calculated using the radius $r$ of the farthest RC data point in the halo of each galaxy.

\begin{figure*}[ht]
\centering
{
\includegraphics[width=0.49\linewidth]{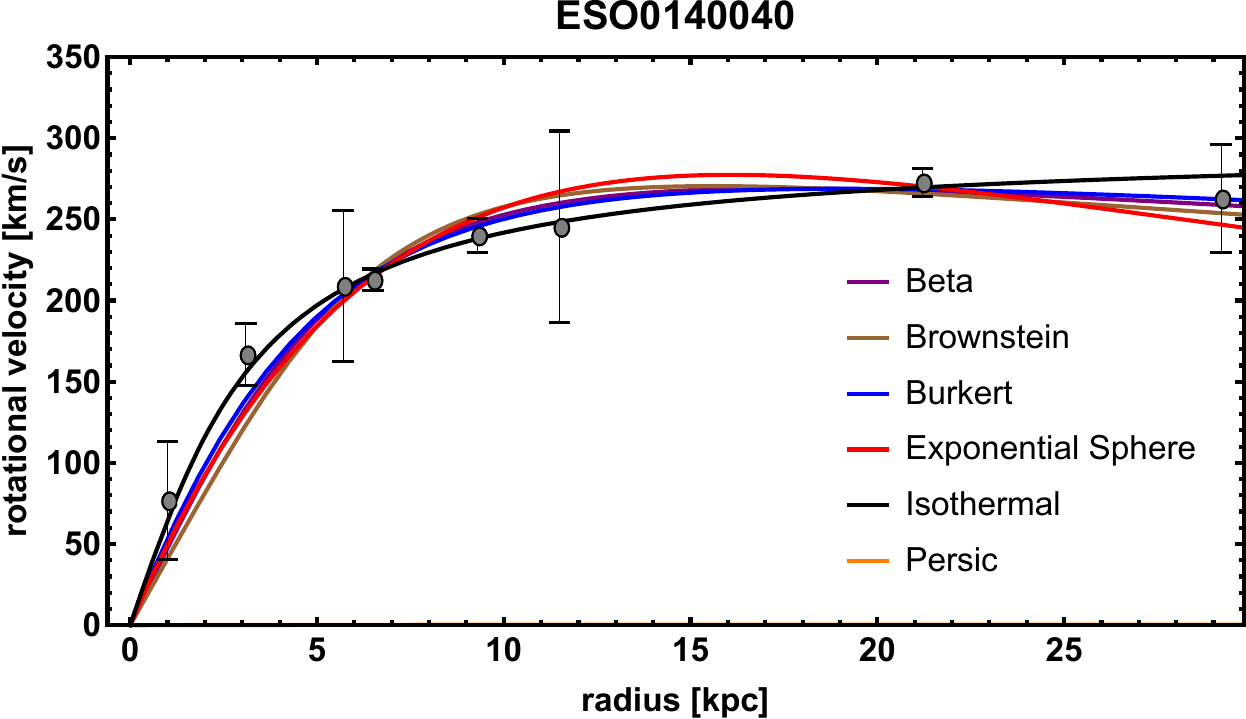}
\hfill
\includegraphics[width=0.49\linewidth]{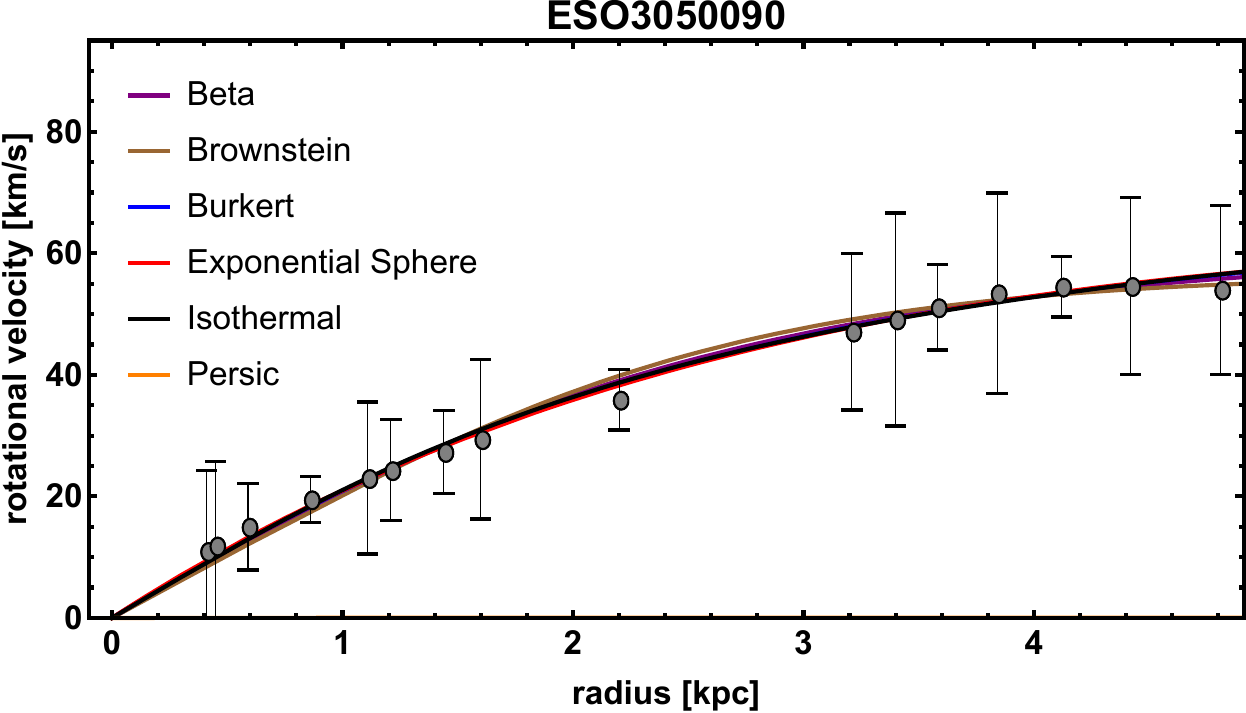}
}\\
\vspace{0.2cm}
{
\includegraphics[width=0.49\linewidth]{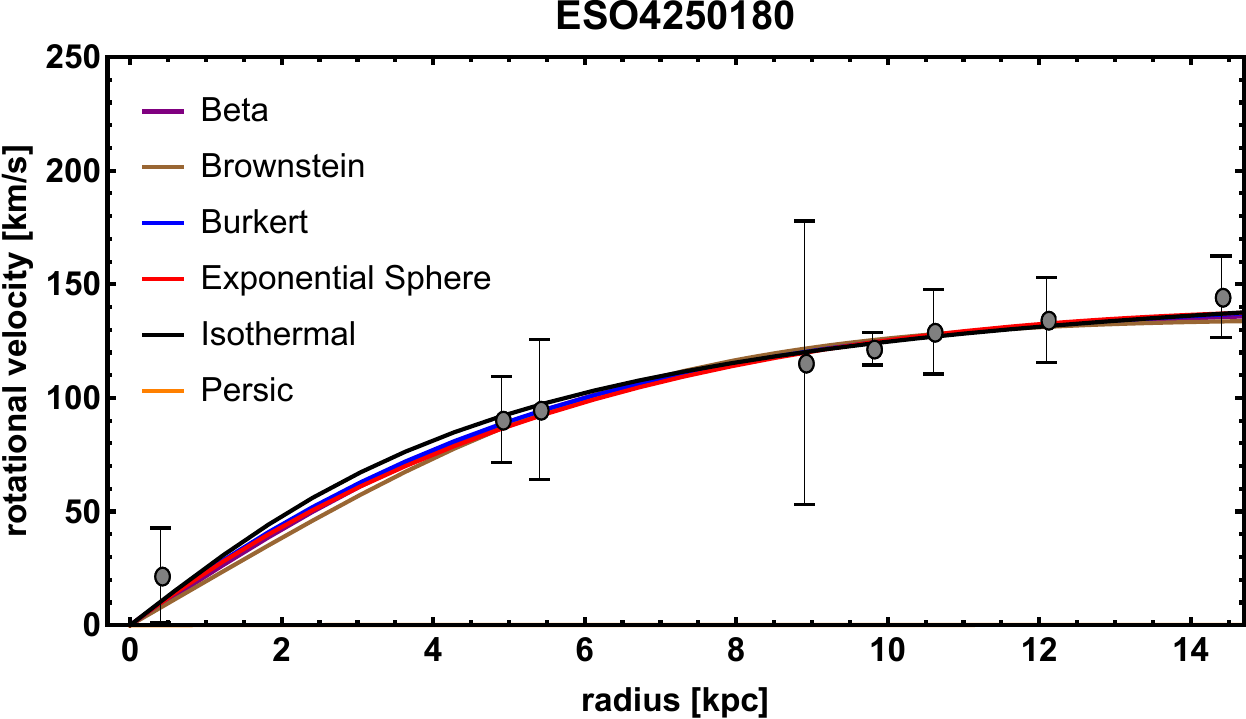}
\hfill
\includegraphics[width=0.49\linewidth]{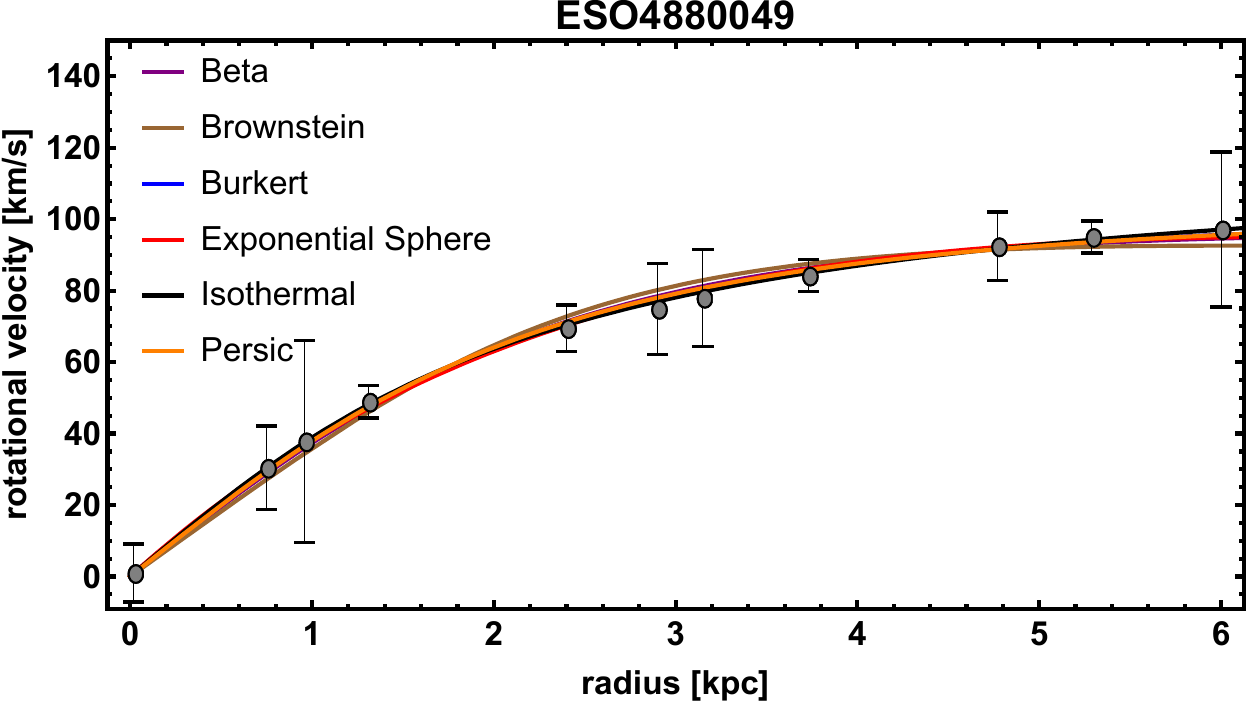}
}\\
\vspace{0.2cm}
{
\includegraphics[width=0.49\linewidth]{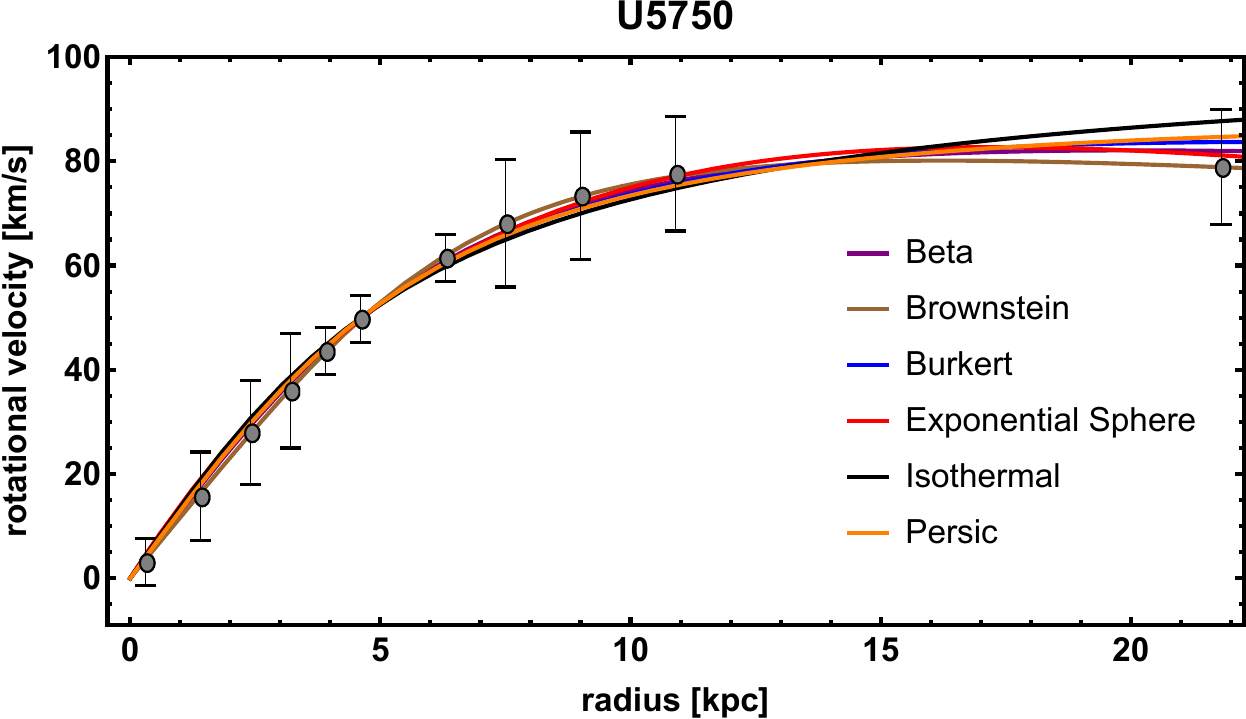}
\hfill
\includegraphics[width=0.49\linewidth]{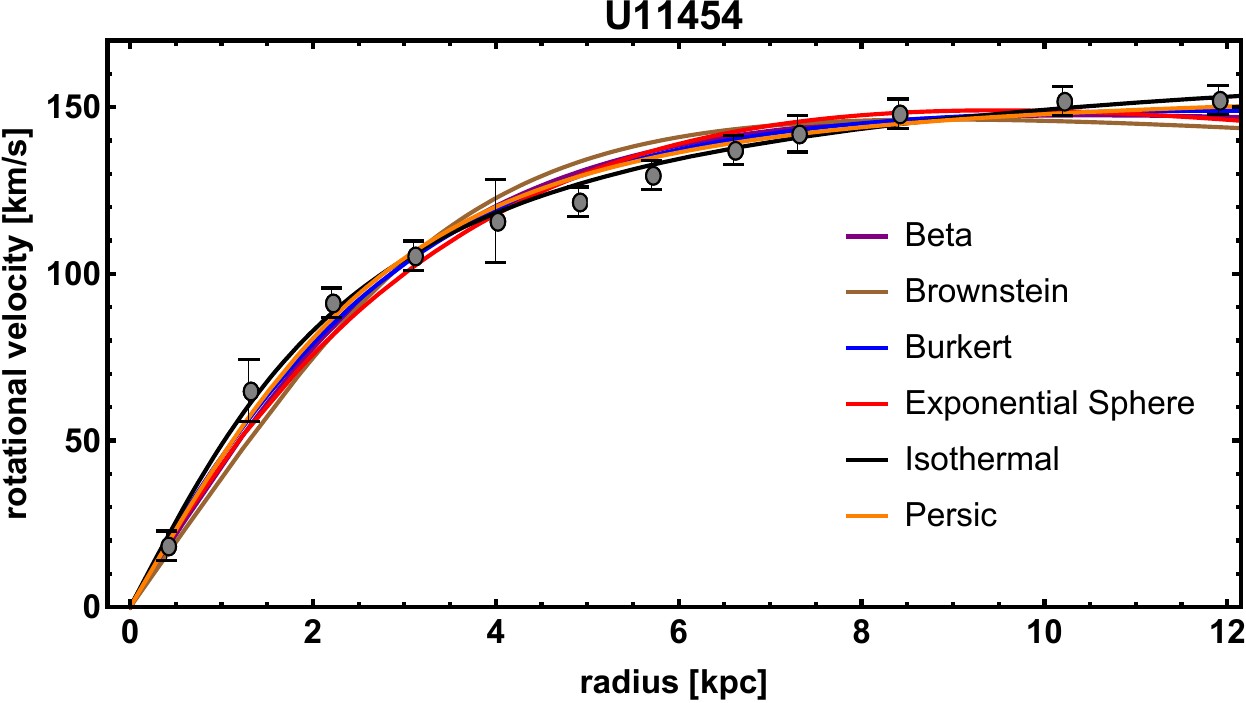}
}\\
\vspace{0.2cm}
{
\includegraphics[width=0.49\linewidth]{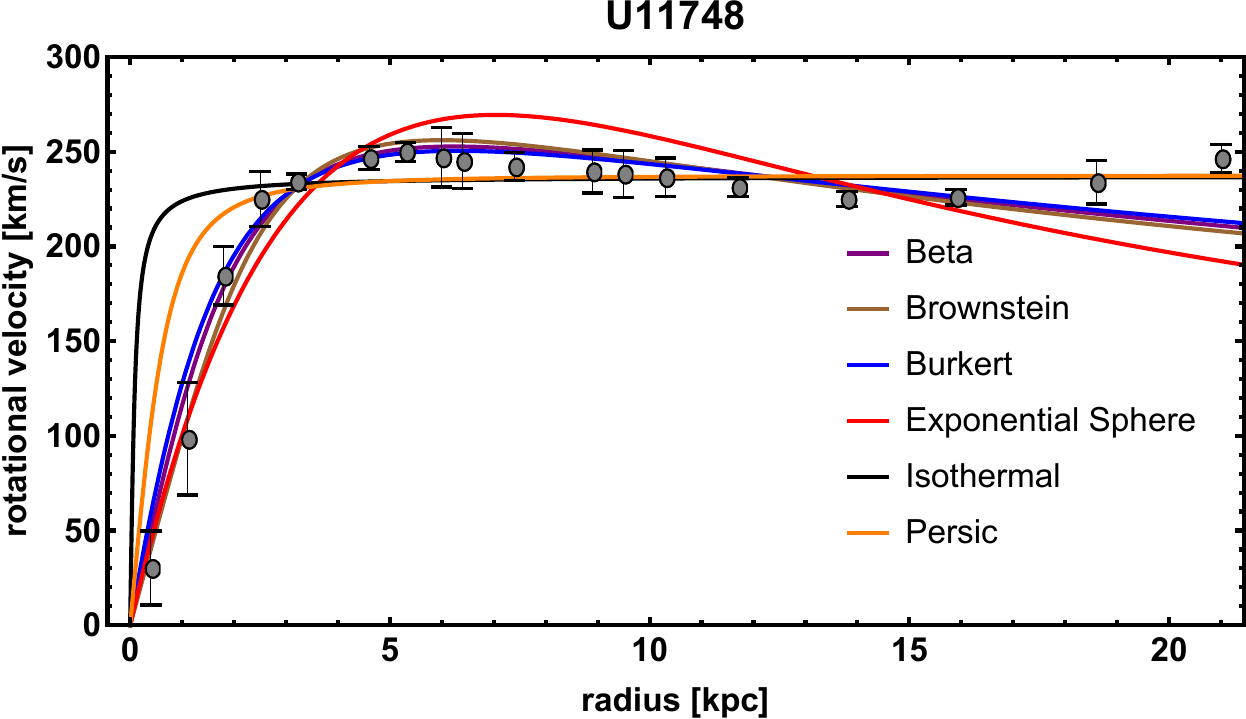}
\hfill
\includegraphics[width=0.49\linewidth]{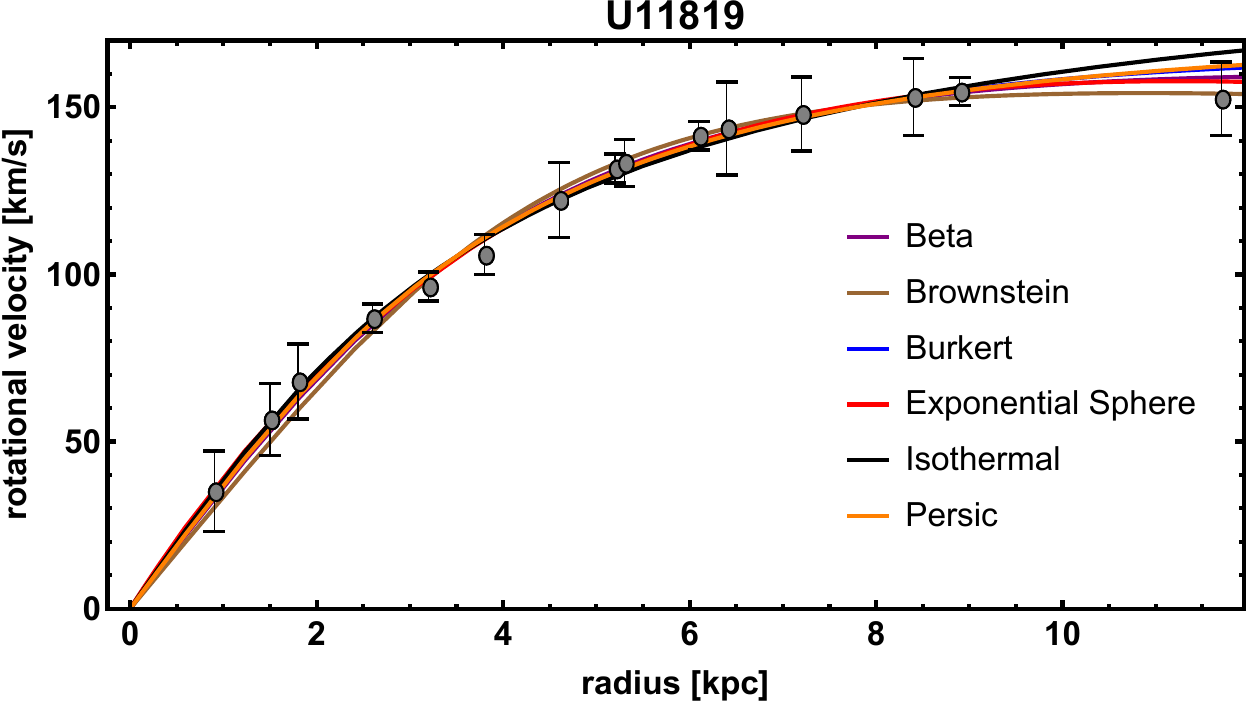}
}\\
\vspace{-0.2cm}
\caption{Observed galaxy RCs fitted using the new phenomenological DM models.}
\label{fig:RC1}
\end{figure*}

\begin{figure*}[ht]
\centering
{
\includegraphics[width=0.49\linewidth]{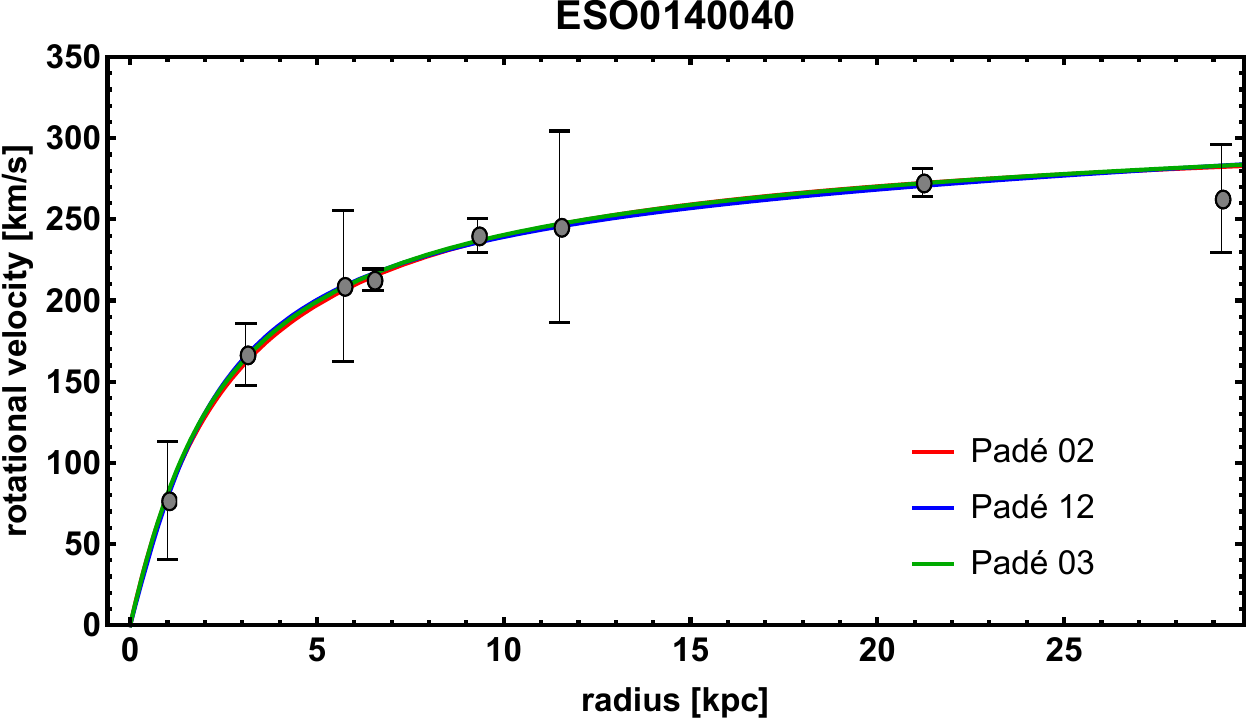}
\hfill
\includegraphics[width=0.49\linewidth]{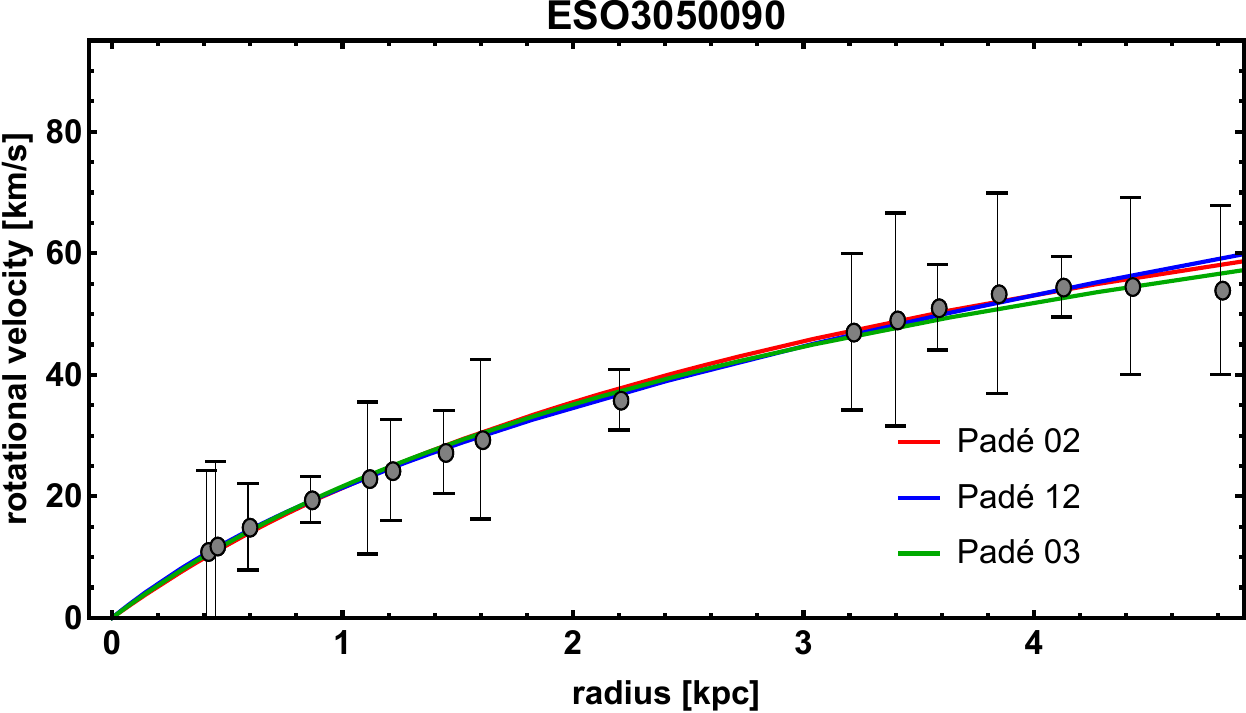}
}\\
\vspace{0.2cm}
{
\includegraphics[width=0.49\linewidth]{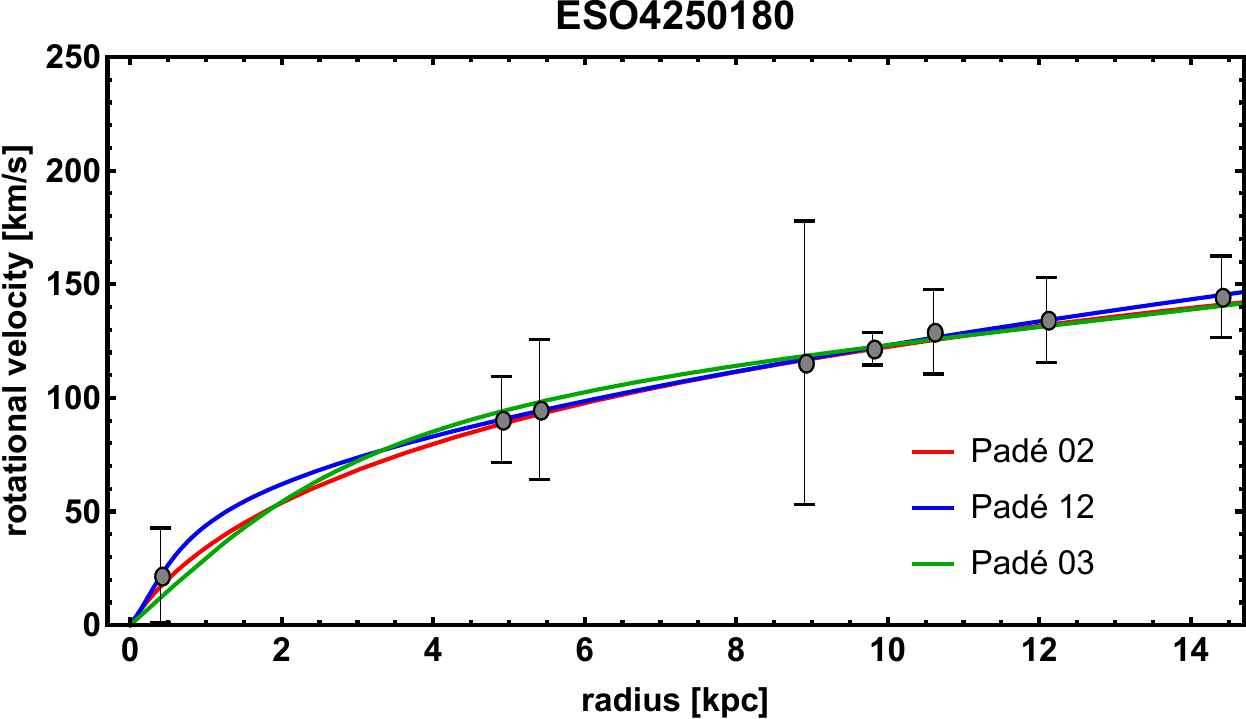}
\hfill
\includegraphics[width=0.49\linewidth]{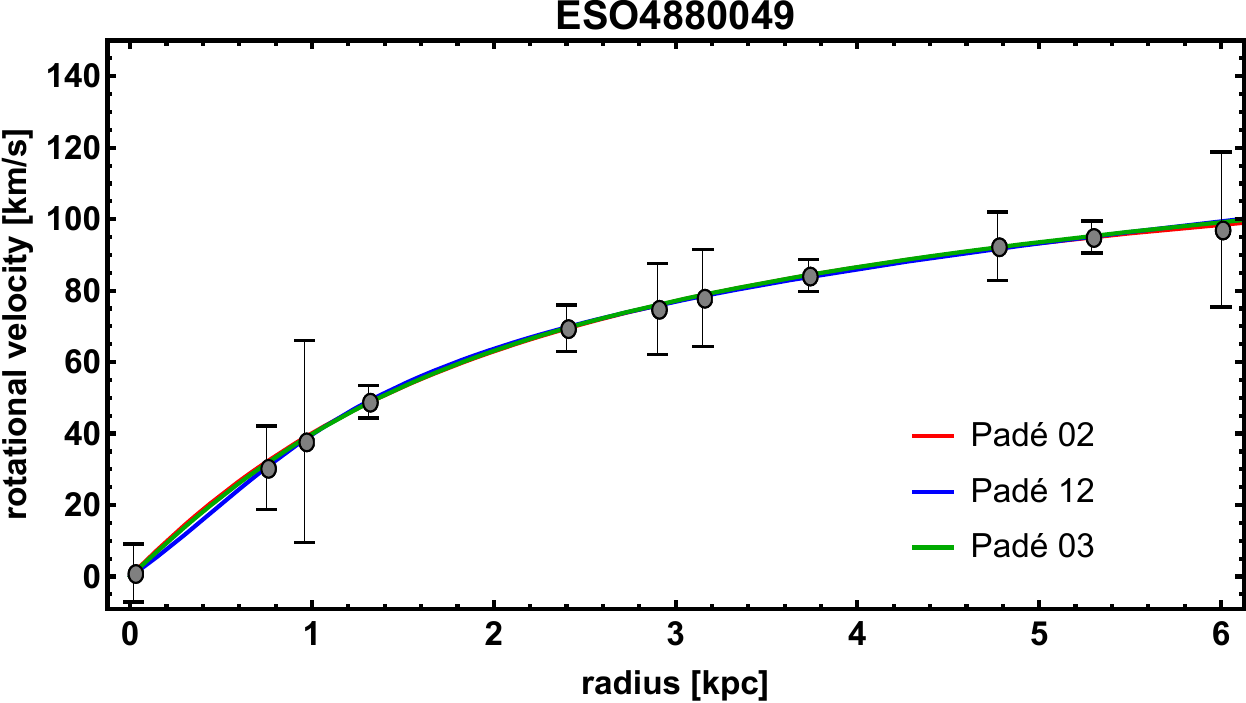}
}\\
\vspace{0.2cm}
{
\includegraphics[width=0.49\linewidth]{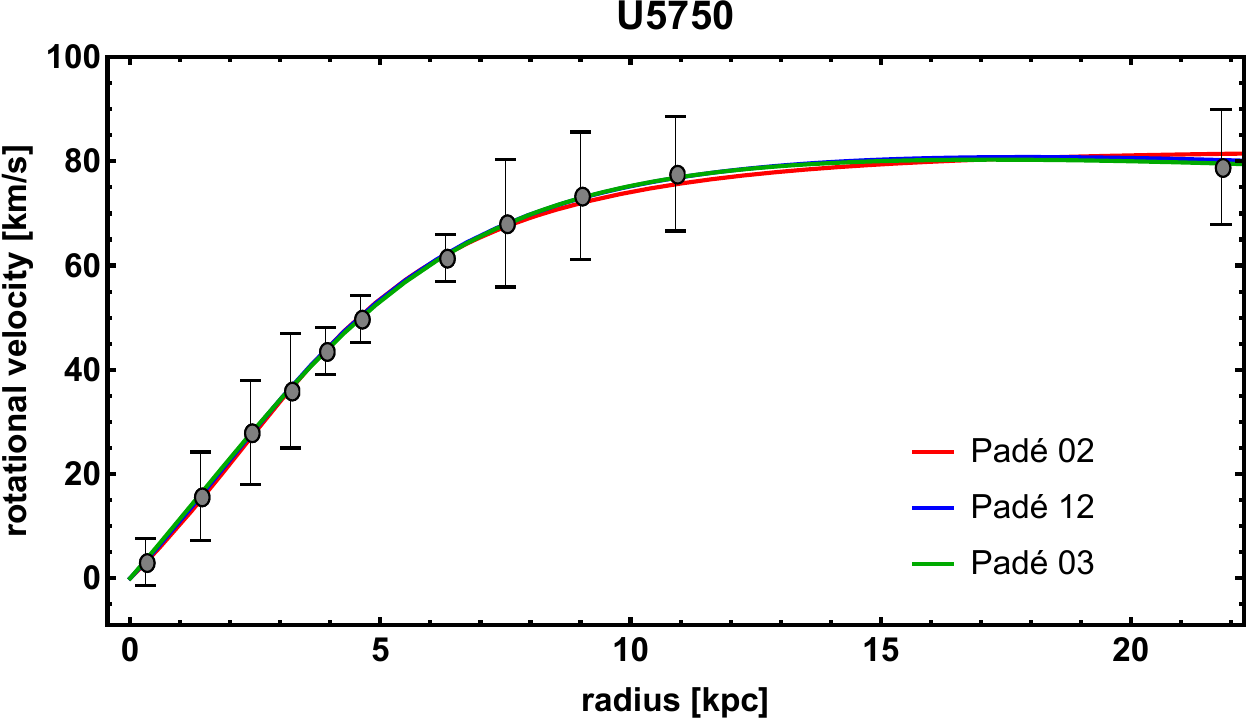}
\hfill
\includegraphics[width=0.49\linewidth]{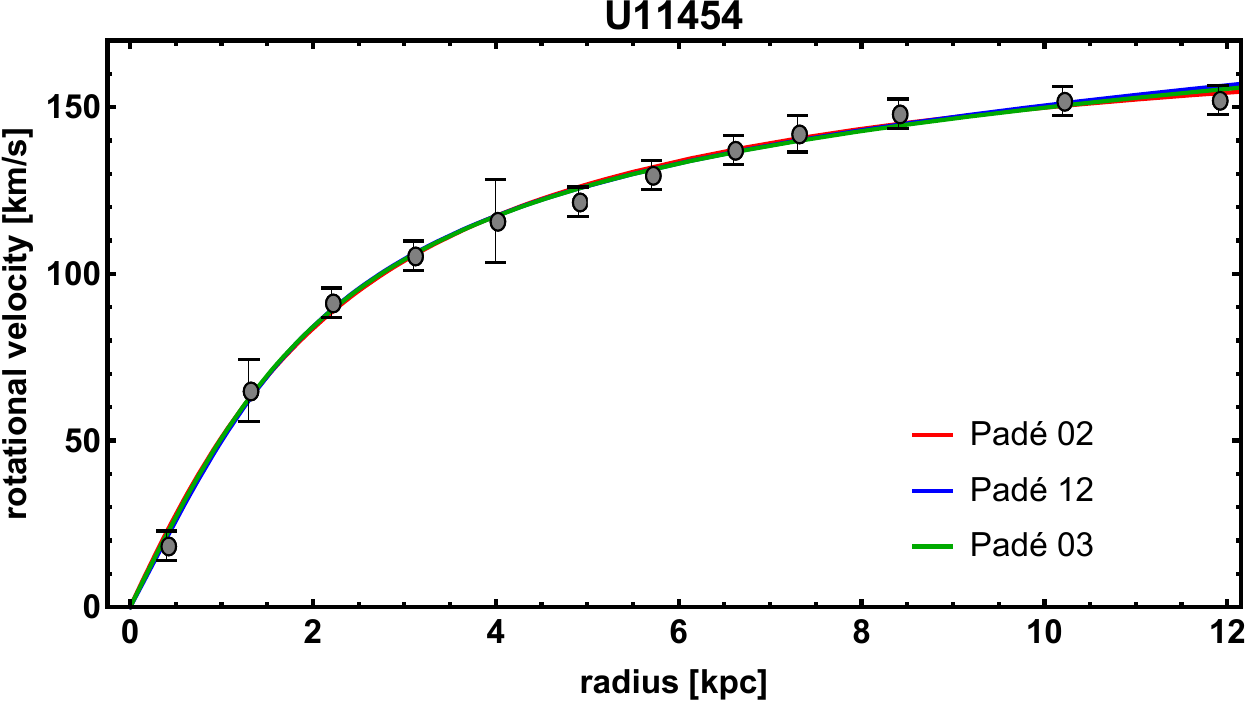}
}\\
\vspace{0.2cm}
{
\includegraphics[width=0.49\linewidth]{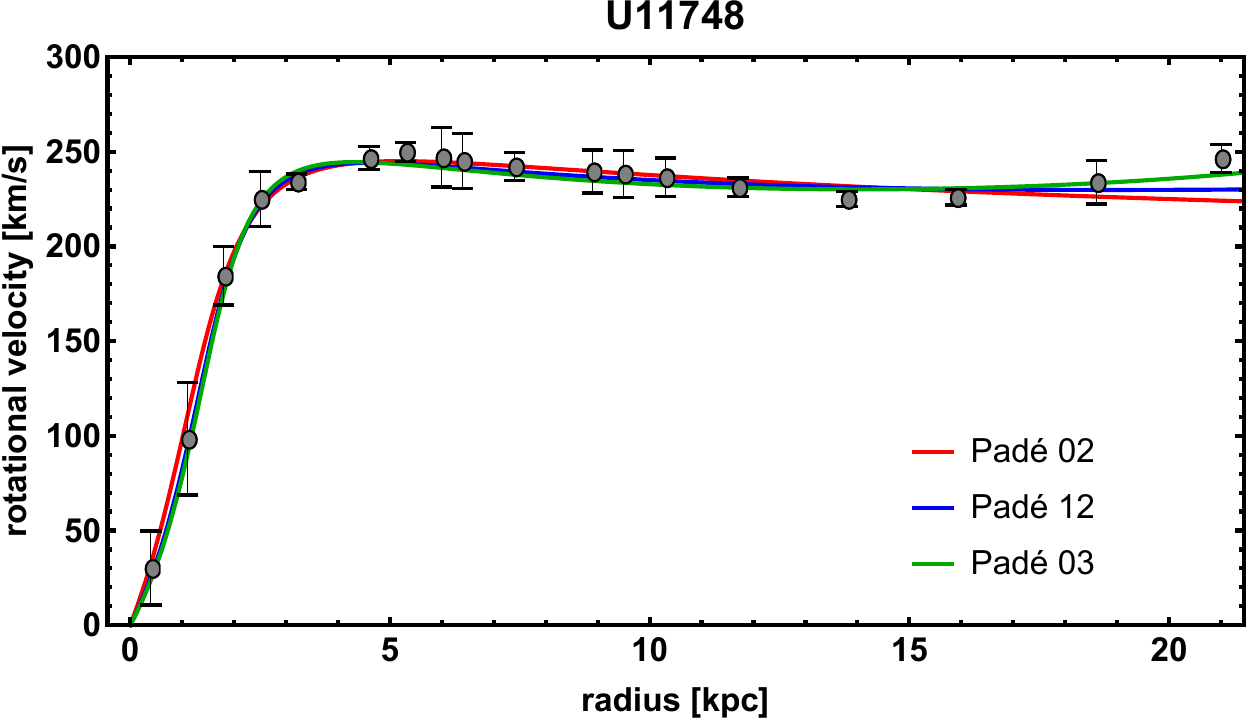}
\hfill
\includegraphics[width=0.49\linewidth]{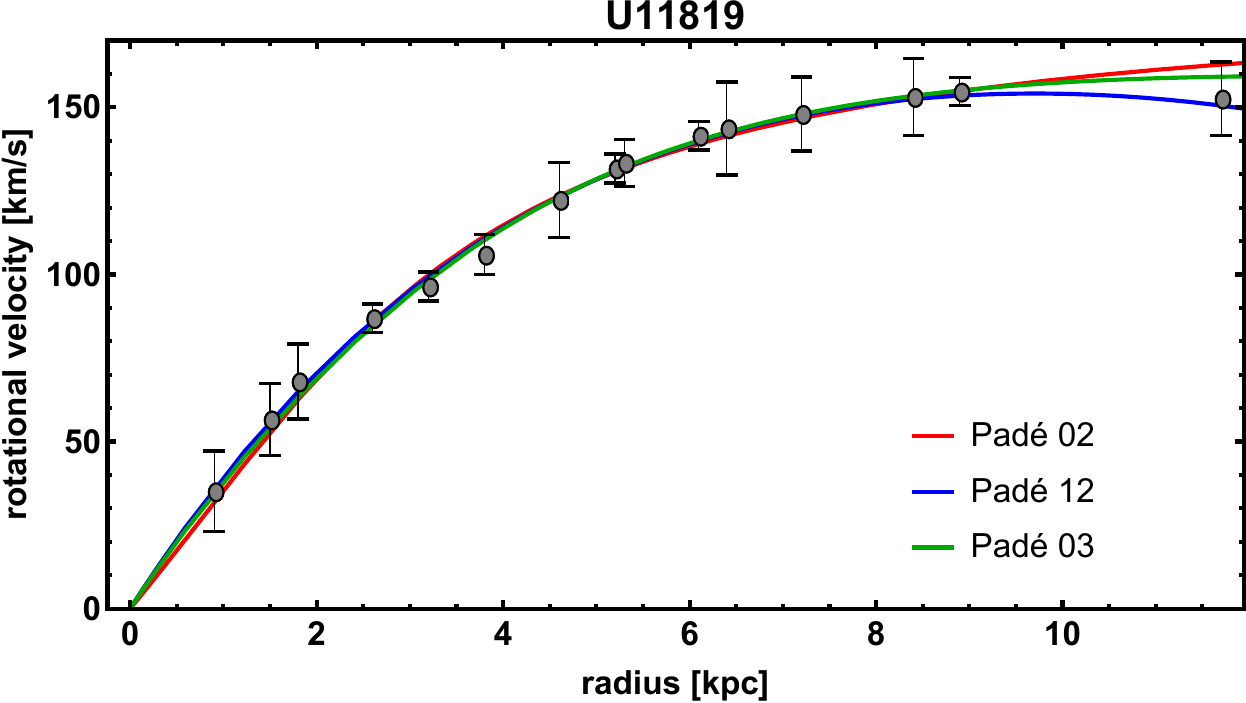}
}\\
\vspace{-0.2cm}
\caption{Observed galaxy RCs fitted using the new phenomenological DM models based on the Pad\'e approximation.}
\label{fig:RC2}
\end{figure*}

The results are summarized in Tables~\ref{tab:ESO0140040}-\ref{tab:U11819} and displayed in Figs.~\ref{fig:RC1}-\ref{fig:RC2}.

\subsection{Results for ESO class of galaxies}

 In this subsection, we present the fitting results for the galaxies ESO0140040, ESO3050090, ESO4250180 and ESO4880049 using both the standard phenomenological DM profiles and the new phenomenological DM models based on the Pad\'e  approximation.
 
Table \ref{tab:ESO0140040} presents the best-fit parameters obtained for the DM standard phenomenological profiles and the the Pad\'e-based  models for the galaxy ESO0140040. 
Among these models, the ISO profile gives the smallest values of the BIC, indicating the best agreement with the observed RC within the set of phenomenological profiles. 
The corresponding DM halo mass is estimated as $M=5.20\times10^{11}M_{\odot}$. 
The remaining phenomenological profiles predict halo masses in the interval $(4.14-4.98)\times10^{11}M_{\odot}$. 
Among the Pad\'e-based DM models, the best performing is the Pad\'e~02 profile, which is weakly excluded ($\Delta{\rm BIC}=2$) and yields a barely constrained halo mass of  $M=5.41\times10^{11}M_{\odot}$. 
The Pad\'e 12 and Pad\'e 03 models are mildly excluded ($\Delta{\rm BIC}=4$) and the corresponding predicted masses are largely undetermined even though the best fit value is about $5.44\times10^{11}M_{\odot}$ for both models. 
Overall, the Pad\'e~02 profile is statistically acceptable and provides a mass which is consistent with the that of the best phenomenological model, namely the ISO profile.

\begin{table}[t]
\footnotesize
\centering
\setlength{\tabcolsep}{.85em}
\renewcommand{\arraystretch}{1.3}
\caption{Best fit model parameters for galaxy ESO0140040.}
\label{tab:ESO0140040}
\begin{tabular}{lccccccccr}
\hline\hline
Profiles                                        & 
$\rho_{0}$\footnote{For the Persic profile, this entry provides $\beta_0$ in units of km/s.}           & 
$r_{0}$                                         &
$a_1$                                           &
$b_1$                                           &
$b_2$                                           &
$M$                                             & 
$-\ln \mathcal L_{\rm m}$                       &
BIC & $\Delta$BIC                               \\
& ($10^{-3}M_\odot/pc^{3}$) & (kpc) & & & &($10^{11}M_{\odot}$) & & & \\
\hline\hline 
Beta                                            & 
$128.80^{+25.04}_{-19.70}$                      &
$5.97^{+0.67}_{-0.61}$                          &
$-$ & $-$ & $-$                                 &
$4.55\pm1.18$                                   &
$3.07$ & $10$ & $5$                             \\
Brownstein                                      &
$93.25^{+16.28}_{-12.10}$                       &
$6.15^{+0.65}_{-0.51}$                          &
$-$ & $-$ & $-$                                 &
$4.37\pm1.43$                                   &
$5.72$ & $16$ & $10$                            \\
Burkert                                         &
$178.43^{+41.49}_{-27.80}$                      &
$5.78^{+0.79}_{-0.59}$                          &
$-$ & $-$ & $-$                                 &
$4.66\pm1.00$                                   &
$1.98$ & $8$ & $3$                              \\
Exp. Sphere                                     &
$162.99^{+24.13}_{-21.76}$                      &
$4.75^{+0.42}_{-0.37}$                          &
$-$ & $-$ & $-$                                 &
$4.14\pm1.38$                                   &
$3.35$ & $11$ & $6$                             \\
ISO                                             &
$250.43^{+134.28}_{-52.41}$                     &
$2.41^{+0.53}_{-0.44}$                          &
$-$ & $-$ & $-$                                 &
$5.20\pm1.08$                                   &
$0.56$ & $5$ & $0$                              \\
Persic                                          &
$275.27^{+14.27}_{-14.14}$                      &
$5.03^{+0.88}_{-0.95}$                          &
$-$ & $-$ & $-$                                 &
$4.98\pm0.83$                                   &
$1.56$ & $7$ & $2$                              \\
Pad\'e~02                                       &
$317.87^{+458.86}_{-131.47}$                    &
$1.32^{+1.85}_{-0.47}$                          &
$-$                                             &
$0.51^{+0.27}_{-1.05}$                          & 
$-$                                             &
$5.41^{+5.73}_{-5.41}$                          &
$0.32$ & $7$ & $2$                              \\
Pad\'e~12                                       &
$122.31^{+158.87}_{-50.45}$                     &
$3.52^{+1.95}_{-1.01}$                          &
$0.34^{+1.84}_{-0.44}$                          &
$-2.67^{+1.83}_{-5.55}$                         & 
$-$                                             &
$5.44^{+23.73}_{-5.44}$                         &
$0.46$ & $9$ & $4$                              \\
Pad\'e~03                                       &
$134.33^{+85.55}_{-35.17}$                      &
$3.89^{+1.49}_{-0.95}$                          &
$-$                                             &
$-0.68^{+1.11}_{-1.35}$                         & 
$1.62^{+1.72}_{-1.29}$                          &
$5.44^{+11.44}_{-5.44}$                         &
$0.39$ & $9$ & $4$                              \\
\hline
\end{tabular}
\end{table}

\begin{table}[t]
\footnotesize
\centering
\setlength{\tabcolsep}{.85em}
\renewcommand{\arraystretch}{1.3}
\caption{Best fit model parameters for galaxy ESO3050090.}
\label{tab:ESO3050090}
\begin{tabular}{lccccccccr}
\hline\hline
Profiles                                        & 
$\rho_{0}$\footnote{For the Persic profile, this entry provides $\beta_0$ in units of km/s.}           & 
$r_{0}$                                         &
$a_1$                                           &
$b_1$                                           &
$b_2$                                           &
$M$                                             & 
$-\ln \mathcal L_{\rm m}$                       &
BIC & $\Delta$BIC                               \\
& ($10^{-3}M_\odot/pc^{3}$) & (kpc) & & & &($10^{9}M_{\odot}$) & & & \\
\hline\hline 
Beta                                            & 
$22.61^{+15.71}_{-5.08}$                        &
$2.84^{+1.69}_{-0.72}$                          &
$-$ & $-$ & $-$                                 &
$3.48\pm0.25$                                   &
$0.42$ & $6$ & $0$                             \\
Brownstein                                      &
$21.17^{+12.21}_{-4.96}$                        &
$2.48^{+1.16}_{-0.54}$                          &
$-$ & $-$ & $-$                                 &
$3.36\pm0.35$                                   &
$0.65$ & $7$ & $1$                            \\
Burkert                                         &
$26.43^{+17.07}_{-6.97}$                        &
$3.27^{+2.64}_{-0.87}$                          &
$-$ & $-$ & $-$                                 &
$3.57\pm0.26$                                   &
$0.30$ & $6$ & $0$                              \\
Exp. Sphere                                     &
$27.84^{+20.25}_{-7.10}$                        &
$2.22^{+1.97}_{-0.58}$                          &
$-$ & $-$ & $-$                                 &
$3.58\pm0.40$                                   &
$0.25$ & $6$ & $0$                             \\
ISO                                             &
$24.11^{+20.11}_{-5.79}$                        &
$1.88^{+1.36}_{-0.51}$                          &
$-$ & $-$ & $-$                                 &
$3.58\pm0.32$                                   &
$0.34$ & $6$ & $0$                              \\
Persic                                          &
$71.40^{+28.03}_{-17.62}$                       &
$3.38^{+2.27}_{-1.39}$                          &
$-$ & $-$ & $-$                                 &
$3.51\pm0.24$                                   &
$0.40$ & $6$ & $0$                              \\
Pad\'e~02                                       &
$22.00^{+20.67}_{-7.39}$                        &
$2.14^{+1.44}_{-0.82}$                          &
$-$                                             &
$0.35^{+0.49}_{-2.07}$                          & 
$-$                                             &
$3.78\pm0.72$                                   &
$0.17$ & $9$ & $3$                              \\
Pad\'e~12                                       &
$22.50^{+21.27}_{-9.79}$                        &
$1.96^{+2.36}_{-1.00}$                          &
$2.09^{+1.01}_{-2.23}$                          &
$0.55^{+3.35}_{-3.72}$                          & 
$-$                                             &
$3.91\pm3.87$                                   &
$0.13$ & $11$ & $5$                              \\
Pad\'e~03                                       &
$19.67^{+12.32}_{-5.87}$                        &
$2.73^{+4.58}_{-0.93}$                          &
$-$                                             &
$1.11^{+0.89}_{-2.28}$                          & 
$2.83^{+2.27}_{-4.72}$                          &
unc                                             &
$0.21$ & $12$ & $6$                              \\
\hline
\end{tabular}
\end{table}

\begin{table}[t]
\footnotesize
\centering
\setlength{\tabcolsep}{.85em}
\renewcommand{\arraystretch}{1.3}
\caption{Best fit model parameters for galaxy ESO4250180.}
\label{tab:ESO4250180}
\begin{tabular}{lccccccccr}
\hline\hline
Profiles                                        & 
$\rho_{0}$\footnote{For the Persic profile, this entry provides $\beta_0$ in units of km/s.}           & 
$r_{0}$                                         &
$a_1$                                           &
$b_1$                                           &
$b_2$                                           &
$M$                                             & 
$-\ln \mathcal L_{\rm m}$                       &
BIC & $\Delta$BIC                               \\
& ($10^{-3}M_\odot/pc^{3}$) & (kpc) & & & &($10^{10}M_{\odot}$) & & & \\
\hline\hline 
Beta                                            & 
$21.37^{+34.29}_{-4.84}$                        &
$5.16^{+4.32}_{-0.91}$                          &
$-$ & $-$ & $-$                                 &
$6.15\pm1.66$                                   &
$0.70$ & $5$ & $0$                             \\
Brownstein                                      &
$18.26^{+29.85}_{-4.59}$                        &
$4.98^{+3.70}_{-0.96}$                          &
$-$ & $-$ & $-$                                 &
$5.98\pm1.97$                                   &
$0.96$ & $6$ & $1$                            \\
Burkert                                         &
$28.08^{+52.46}_{-6.10}$                        &
$4.85^{+4.94}_{-0.74}$                          &
$-$ & $-$ & $-$                                 &
$6.25\pm1.57$                                   &
$0.54$ & $5$ & $0$                              \\
Exp. Sphere                                     &
$27.48^{+35.74}_{-7.30}$                        &
$4.38^{+3.66}_{-0.90}$                          &
$-$ & $-$ & $-$                                 &
$6.29\pm2.00$                                   &
$0.55$ & $5$ & $0$                             \\
ISO                                             &
$19.96^{+48.57}_{-4.93}$                        &
$3.14^{+4.45}_{-0.73}$                          &
$-$ & $-$ & $-$                                 &
$6.29\pm2.65$                                   &
$0.47$ & $5$ & $0$                              \\
Persic                                          &
$151.89^{+64.82}_{-27.53}$                      &
$7.50^{+6.64}_{-5.14}$                          &
$-$ & $-$ & $-$                                 &
$6.19\pm1.06$                                   &
$0.61$ & $5$ & $0$                              \\
Pad\'e~02                                       &
$20.53^{+52.63}_{-9.33}$                        &
$3.27^{+5.30}_{-0.80}$                          &
$-$                                             &
$0.09^{+1.04}_{-2.89}$                          & 
$-$                                             &
$6.67^{+36.84}_{-6.67}$                         &
$0.09$ & $6$ & $1$                              \\
Pad\'e~12                                       &
$39.36^{+109.96}_{-15.45}$                      &
$1.45^{+3.23}_{-0.51}$                          &
$1.73^{+1.34}_{-1.41}$                          &
$-1.36^{+3.47}_{-2.24}$                         & 
$-$                                             &
$7.06\pm2.54$                                   &
$0.01$ & $8$ & $3$                              \\
Pad\'e~03                                       &
$14.54^{+23.80}_{-5.87}$                        &
$7.79^{+5.65}_{-1.47}$                          &
$-$                                             &
$0.36^{+2.52}_{-3.14}$                          & 
$-0.00^{+5.81}_{-5.50}$                         &
$6.61^{+15.75}_{-6.61}$                         &
$0.27$ & $9$ & $4$                             \\
\hline
\end{tabular}
\end{table}

\begin{table}[t]
\footnotesize
\centering
\setlength{\tabcolsep}{.85em}
\renewcommand{\arraystretch}{1.3}
\caption{Best fit model parameters for galaxy ESO4880049.}
\label{tab:ESO4880049}
\begin{tabular}{lccccccccr}
\hline\hline
Profiles                                        & 
$\rho_{0}$\footnote{For the Persic profile, this entry provides $\beta_0$ in units of km/s.}           & 
$r_{0}$                                         &
$a_1$                                           &
$b_1$                                           &
$b_2$                                           &
$M$                                             & 
$-\ln \mathcal L_{\rm m}$                       &
BIC & $\Delta$BIC                               \\
& ($10^{-3}M_\odot/pc^{3}$) & (kpc) & & & &($10^{10}M_{\odot}$) & & & \\
\hline\hline 
Beta                                            & 
$85.59^{+31.32}_{-17.51}$                       &
$2.53^{+0.56}_{-0.41}$                          &
$-$ & $-$ & $-$                                 &
$1.25\pm0.06$                                   &
$0.40$ & $6$ & $1$                             \\
Brownstein                                      &
$73.79^{+22.33}_{-15.53}$                       &
$2.37^{+0.43}_{-0.33}$                          &
$-$ & $-$ & $-$                                 &
$1.20\pm0.06$                                   &
$1.02$ & $7$ & $2$                            \\
Burkert                                         &
$107.94^{+40.85}_{-24.82}$                      &
$2.64^{+0.77}_{-0.42}$                          &
$-$ & $-$ & $-$                                 &
$1.28\pm0.07$                                   &
$0.21$ & $5$ & $0$                              \\
Exp. Sphere                                     &
$115.57^{+37.72}_{-25.78}$                      &
$1.85^{+0.50}_{-0.25}$                          &
$-$ & $-$ & $-$                                 &
$1.26\pm0.07$                                   &
$0.27$ & $5$ & $0$                             \\
ISO                                             &
$99.06^{+39.94}_{-24.10}$                       &
$1.55^{+0.52}_{-0.26}$                          &
$-$ & $-$ & $-$                                 &
$1.31\pm0.07$                                   &
$0.08$ & $5$ & $0$                              \\
Persic                                          &
$105.33^{+9.55}_{-9.66}$                        &
$2.59^{+0.65}_{-0.57}$                          &
$-$ & $-$ & $-$                                 &
$1.27\pm0.05$                                   &
$0.25$ & $5$ & $0$                              \\
Pad\'e~02                                       &
$86.31^{+64.22}_{-32.58}$                       &
$1.61^{+1.15}_{-0.54}$                          &
$-$                                             &
$0.12^{+0.62}_{-1.84}$                          & 
$-$                                             &
$1.35\pm0.08$                                   &
$0.02$ & $7$ & $2$                              \\
Pad\'e~12                                       &
$54.77^{+58.60}_{-19.60}$                       &
$1.91^{+1.54}_{-0.61}$                          &
$1.58^{+1.51}_{-1.00}$                          &
$-1.37^{+1.79}_{-2.91}$                         & 
$-$                                             &
$1.38\pm0.06$                                   &
$0.02$ & $10$ & $5$                              \\
Pad\'e~03                                       &
$65.07^{+36.30}_{-23.31}$                       &
$2.29^{+0.93}_{-0.50}$                          &
$-$                                             &
$-0.68^{+2.29}_{-2.33}$                         & 
$0.82^{+4.71}_{-3.55}$                          &
$1.37\pm0.68$                                   &
$0.02$ & $10$ & $5$                             \\
\hline
\end{tabular}
\end{table}

\begin{table}[t]
\footnotesize
\centering
\setlength{\tabcolsep}{.85em}
\renewcommand{\arraystretch}{1.3}
\caption{Best fit model parameters for galaxy U5750.}
\label{tab:U5750}
\begin{tabular}{lccccccccr}
\hline\hline
Profiles                                        & 
$\rho_{0}$\footnote{For the Persic profile, this entry provides $\beta_0$ in units of km/s.}           & 
$r_{0}$                                         &
$a_1$                                           &
$b_1$                                           &
$b_2$                                           &
$M$                                             & 
$-\ln \mathcal L_{\rm m}$                       &
BIC & $\Delta$BIC                               \\
& ($10^{-3}M_\odot/pc^{3}$) & (kpc) & & & &($10^{10}M_{\odot}$) & & & \\
\hline\hline 
Beta                                            & 
$8.89^{+3.25}_{-1.75}$                          &
$6.80^{+2.00}_{-1.49}$                          &
$-$ & $-$ & $-$                                 &
$3.41\pm0.36$                                   &
$0.13$ & $5$ & $0$                             \\
Brownstein                                      &
$7.32^{+2.25}_{-1.20}$                          &
$6.29^{+1.75}_{-1.07}$                          &
$-$ & $-$ & $-$                                 &
$3.15\pm0.33$                                   &
$0.02$ & $5$ & $0$                            \\
Burkert                                         &
$11.22^{+4.75}_{-2.34}$                         &
$6.99^{+2.47}_{-1.59}$                          &
$-$ & $-$ & $-$                                 &
$3.55\pm0.34$                                   &
$0.32$ & $6$ & $1$                              \\
Exp. Sphere                                     &
$11.72^{+4.36}_{-2.16}$                         &
$5.02^{+1.74}_{-0.96}$                          &
$-$ & $-$ & $-$                                 &
$3.34\pm0.31$                                   &
$0.21$ & $5$ & $0$                             \\
ISO                                             &
$10.01^{+5.67}_{-2.14}$                         &
$4.13^{+1.73}_{-1.12}$                          &
$-$ & $-$ & $-$                                 &
$3.90\pm0.41$                                   &
$0.64$ & $6$ & $1$                              \\
Persic                                          &
$88.84^{+18.80}_{-12.71}$                       &
$7.06^{+2.13}_{-2.01}$                          &
$-$ & $-$ & $-$                                 &
$3.63\pm0.37$                                   &
$0.30$ & $6$ & $1$                              \\
Pad\'e~02                                       &
$6.30^{+7.33}_{-2.44}$                          &
$5.55^{+3.35}_{-2.24}$                          &
$-$                                             &
$-0.57^{+0.90}_{-2.33}$                         & 
$-$                                             &
$3.36\pm0.37$                                   &
$0.08$ & $8$ & $3$                              \\
Pad\'e~12                                       &
$6.05^{+7.14}_{-2.15}$                          &
$5.51^{+3.69}_{-2.35}$                          &
$0.48^{+2.30}_{-0.92}$                          &
$-1.50^{+1.71}_{-1.50}$                         & 
$-$                                             &
$3.26\pm0.57$                                   &
$0.03$ & $10$ & $5$                              \\
Pad\'e~03                                       &
$7.48^{+4.32}_{-3.31}$                          &
$6.41^{+2.57}_{-2.03}$                          &
$-$                                             &
$0.53^{+1.89}_{-2.81}$                          & 
$-1.30^{+4.09}_{-2.70}$                         &
$3.21\pm0.41$                                   &
$0.02$ & $10$ & $5$                             \\
\hline
\end{tabular}
\end{table}

\begin{table}[t]
\footnotesize
\centering
\setlength{\tabcolsep}{.85em}
\renewcommand{\arraystretch}{1.3}
\caption{Best fit model parameters for galaxy U11454.}
\label{tab:U11454}
\begin{tabular}{lccccccccr}
\hline\hline
Profiles                                        & 
$\rho_{0}$\footnote{For the Persic profile, this entry provides $\beta_0$ in units of km/s.}           & 
$r_{0}$                                         &
$a_1$                                           &
$b_1$                                           &
$b_2$                                           &
$M$                                             & 
$-\ln \mathcal L_{\rm m}$                       &
BIC & $\Delta$BIC                               \\
& ($10^{-3}M_\odot/pc^{3}$) & (kpc) & & & &($10^{10}M_{\odot}$) & & & \\
\hline\hline 
Beta                                            & 
$106.93^{+13.61}_{-11.73}$                      &
$3.60^{+0.30}_{-0.22}$                          &
$-$ & $-$ & $-$                                 &
$5.99\pm0.19$                                   &
$7.57$ & $20$ & $11$                             \\
Brownstein                                      &
$84.24^{+10.04}_{-8.13}$                        &
$3.54^{+0.22}_{-0.22}$                          &
$-$ & $-$ & $-$                                 &
$5.74\pm0.24$                                   &
$14.63$ & $34$ & $25$                              \\
Burkert                                         &
$140.29^{+21.00}_{-15.15}$                      &
$3.64^{+0.32}_{-0.26}$                          &
$-$ & $-$ & $-$                                 &
$6.14\pm0.16$                                   &
$4.98$ & $15$ & $6$                              \\
Exp. Sphere                                     &
$133.12^{+16.37}_{-11.88}$                      &
$2.79^{+0.20}_{-0.15}$                          &
$-$ & $-$ & $-$                                 &
$5.93\pm0.22$                                   &
$8.87$ & $23$ & $14$                             \\
ISO                                             &
$151.02^{+29.62}_{-20.37}$                      &
$1.90^{+0.22}_{-0.18}$                          &
$-$ & $-$ & $-$                                 &
$6.48\pm0.10$                                   &
$1.89$ & $9$ & $0$                              \\
Persic                                          &
$155.94^{+4.44}_{-4.45}$                        &
$3.32^{+0.33}_{-0.34}$                          &
$-$ & $-$ & $-$                                 &
$6.23\pm0.13$                                   &
$4.05$ & $13$ & $4$                              \\
Pad\'e~02                                       &
$152.08^{+55.81}_{-39.71}$                      &
$1.71^{+0.84}_{-0.39}$                          &
$-$                                             &
$0.12^{+0.41}_{-0.66}$                          & 
$-$                                             &
$6.58\pm0.26$                                   &
$1.69$ & $11$ & $2$                              \\
Pad\'e~12                                       &
$83.22^{+62.75}_{-21.67}$                       &
$2.46^{+1.12}_{-0.48}$                          &
$0.46^{+1.45}_{-0.46}$                          &
$-1.86^{+1.43}_{-2.22}$                         & 
$-$                                             &
$6.76\pm1.91$                                   &
$1.59$ & $13$ & $4$                             \\
Pad\'e~03                                       &
$106.01^{+36.62}_{-25.21}$                      &
$2.53^{+0.71}_{-0.46}$                          &
$-$                                             &
$-0.90^{+0.87}_{-1.16}$                         & 
$1.98^{+1.54}_{-1.03}$                          &
$6.67\pm1.29$                                   &
$1.63$ & $13$ & $4$                             \\
\hline
\end{tabular}
\end{table}

\begin{table}[t]
\footnotesize
\centering
\setlength{\tabcolsep}{.85em}
\renewcommand{\arraystretch}{1.3}
\caption{Best fit model parameters for galaxy U11748.}
\label{tab:U11748}
\begin{tabular}{lccccccccr}
\hline\hline
Profiles                                        & 
$\rho_{0}$\footnote{For the Persic profile, this entry provides $\beta_0$ in units of km/s.}           & 
$r_{0}$                                         &
$a_1$                                           &
$b_1$                                           &
$b_2$                                           &
$M$                                             & 
$-\ln \mathcal L_{\rm m}$                       &
BIC & $\Delta$BIC                               \\
& ($10^{-3}M_\odot/pc^{3}$) & (kpc) & & & &($10^{11}M_{\odot}$) & & & \\
\hline\hline 
Beta                                            & 
$900.78^{+91.33}_{-82.59}$                      &
$2.13^{+0.10}_{-0.10}$                          &
$-$ & $-$ & $-$                                 &
$2.17\pm0.07$                                   &
$20.72$ & $47$ & $26$                            \\
Brownstein                                      &
$568.34^{+51.90}_{-45.21}$                      &
$2.38^{+0.11}_{-0.09}$                          &
$-$ & $-$ & $-$                                 &
$2.11\pm0.07$                                   &
$25.60$ & $57$ & $36$                           \\
Burkert                                         &
$1441.50^{+156.29}_{-152.23}$                   &
$1.94^{+0.11}_{-0.10}$                          &
$-$ & $-$ & $-$                                 &
$2.22\pm0.09$                                   &
$18.14$ & $42$ & $21$                           \\
Exp. Sphere                                     &
$799.53^{+54.04}_{-50.34}$                      &
$2.08^{+0.06}_{-0.06}$                          &
$-$ & $-$ & $-$                                 &
$1.80\pm0.11$                                   &
$74.53$ & $155$ & $134$                            \\
ISO                                             &
$47314.21^{+68738.50}_{-10548.83}$              &
$0.13^{+0.04}_{-0.03}$                          &
$-$ & $-$ & $-$                                 &
$2.73^{+11.81}_{-2.73}$                         &
$28.07$ & $62$ & $41$                              \\
Persic                                          &
$237.14^{+3.41}_{-2.68}$                        &
$0.77^{+0.27}_{-0.29}$                          &
$-$ & $-$ & $-$                                 &
$2.75\pm0.16$                                   &
$23.99$ & $54$ & $32$                             \\
Pad\'e~02                                       &
$287.10^{+111.05}_{-40.89}$                     &
$2.99^{+0.39}_{-0.31}$                          &
$-$                                             &
$-2.98^{+0.30}_{-0.28}$                         & 
$-$                                             &
$2.45\pm0.15$                                   &
$8.07$ & $25$ & $3$                             \\
Pad\'e~12                                       &
$180.43^{+103.90}_{-44.11}$                     &
$3.56^{+0.55}_{-0.48}$                          &
$0.08^{+0.19}_{-0.08}$                          &
$-3.66^{+0.54}_{-0.63}$                         & 
$-$                                             &
$2.58\pm0.32$                                   &
$6.27$ & $24$ & $3$                             \\
Pad\'e~03                                       &
$168.38^{+92.17}_{-33.02}$                      &
$3.55^{+0.59}_{-0.34}$                          &
$-$                                             &
$-3.75^{+0.44}_{-0.52}$                         & 
$5.07^{+0.97}_{-0.61}$                          &
$2.76\pm0.17$                                   &
$4.88$ & $21$ & $0$                             \\
\hline
\end{tabular}
\end{table}

\begin{table}[t]
\footnotesize
\centering
\setlength{\tabcolsep}{.85em}
\renewcommand{\arraystretch}{1.3}
\caption{Best fit model parameters for galaxy U11819.}
\label{tab:U11819}
\begin{tabular}{lccccccccr}
\hline\hline
Profiles                                        & 
$\rho_{0}$\footnote{For the Persic profile, this entry provides $\beta_0$ in units of km/s.}           & 
$r_{0}$                                         &
$a_1$                                           &
$b_1$                                           &
$b_2$                                           &
$M$                                             & 
$-\ln \mathcal L_{\rm m}$                       &
BIC & $\Delta$BIC                               \\
& ($10^{-3}M_\odot/pc^{3}$) & (kpc) & & & &($10^{10}M_{\odot}$) & & & \\
\hline\hline 
Beta                                            & 
$76.42^{+9.84}_{-8.03}$                         &
$4.61^{+0.43}_{-0.41}$                          &
$-$ & $-$ & $-$                                 &
$6.87\pm0.14$                                   &
$1.09$ & $8$ & $1$                           \\
Brownstein                                      &
$62.52^{+6.98}_{-5.58}$                         &
$4.31^{+0.35}_{-0.29}$                          &
$-$ & $-$ & $-$                                 &
$6.46\pm0.16$                                   &
$1.80$ & $9$ & $2$                           \\
Burkert                                         &
$97.04^{+15.18}_{-10.30}$                       &
$4.78^{+0.59}_{-0.41}$                          &
$-$ & $-$ & $-$                                 &
$7.10\pm0.16$                                   &
$1.20$ & $8$ & $1$                           \\
Exp. Sphere                                     &
$107.41^{+13.81}_{-12.09}$                      &
$3.30^{+0.37}_{-0.26}$                          &
$-$ & $-$ & $-$                                 &
$6.77\pm0.13$                                   &
$0.81$ & $7$ & $0$                         \\
ISO                                             &
$87.05^{+16.53}_{-9.68}$                        &
$2.90^{+0.41}_{-0.30}$                          &
$-$ & $-$ & $-$                                 &
$7.53\pm0.21$                                   &
$1.98$ & $9$ & $2$                           \\ 
Persic                                          &
$174.08^{+9.17}_{-7.37}$                        &
$4.59^{+0.54}_{-0.45}$                          &
$-$ & $-$ & $-$                                 &
$7.16\pm0.16$                                   &
$1.36$ & $8$ & $1$                           \\
Pad\'e~02                                       &
$68.28^{+25.11}_{-18.21}$                       &
$3.66^{+1.14}_{-0.86}$                          &
$-$                                             &
$-0.54^{+0.76}_{-1.07}$                         & 
$-$                                             &
$7.21\pm0.40$                                   &
$1.64$ & $11$ & $4$                             \\
Pad\'e~12                                       &
$50.76^{+33.25}_{-15.63}$                       &
$4.11^{+1.67}_{-1.09}$                          &
$0.02^{+2.52}_{-0.41}$                          &
$-1.06^{+1.01}_{-2.09}$                         & 
$-$                                             &
$6.16\pm0.65$                                   &
$0.77$ & $12$ & $5$                             \\
Pad\'e~03                                       &
$61.24^{+22.35}_{-15.32}$                      &
$4.19^{+0.92}_{-0.93}$                          &
$-$                                             &
$-0.01^{+0.90}_{-1.73}$                         & 
$0.15^{+2.61}_{-1.13}$                          &
$6.89\pm1.87$                                   &
$0.89$ & $13$ & $6$                             \\
\hline
\end{tabular}
\end{table}

Table~\ref{tab:ESO3050090} summarizes the results obtained for the DM phenomenological profiles and the Pad\'e-based ones in the case of the galaxy ESO3050090. 
Several profiles are equally suited to describe the observed RC ($\Delta{\rm BIC}=0$), the Brownstein and the Pad\'e~02 profiles are weakly excluded ($\Delta{\rm BIC}\leq3$), whereas Pad\'e~12 and Pad\'e~03 are mildly excluded ($3<\Delta{\rm BIC}\leq6$). 
All models point to a value of the halo total mass of $M=(3.36-3.91)\times10^{9}M_{\odot}$, with the exception of the Pad\'e~03 model for which the mass is largely unconstrained.

Table~\ref{tab:ESO4250180} lists the results obtained for the DM standard phenomenological profiles and the Pad\'e models for the galaxy ESO4250180. 
Like in the previous case, several models equally describe the observed RC ($\Delta{\rm BIC}=0$), while the Brownstein, the Pad\'e~02 and the Pad\'e~12
profiles are weakly excluded ($\Delta{\rm BIC}\leq3$), and the Pad\'e~03 is mildly excluded ($3<\Delta{\rm BIC}\leq6$). 
The halo masses predicted by all models lie in the range $M=(5.98-7.06)\times10^{10}M_{\odot}$, although all Pad\'e models have huge attached error bars.

The best-fit parameters of the DM phenomenological profiles and Pad\'e approximations for the galaxy ESO4880049 are displayed in Table~\ref{tab:ESO4880049}. 
Here, Burkert, Exponential Sphere, ISO, and Persic models give the best statistical agreement with the observed RC ($\Delta\mathrm{BIC}=0$), then Brownstein, Pad\'e~02 models also provide acceptable fits ($\Delta\mathrm{BIC}\leq3$), and finally Pad\'e~12 and 03 are mildly excluded ($3<\Delta\mathrm{BIC}\leq6$). 
All models provide but Pad\'e~03 provide precisely constrained halo masses in the narrow range $M=(1.20-1.38)\times10^{10}M_{\odot}$. All Pad\'e profiles lead to noticeably larger halo masses.


\subsection{Results for the U class of galaxies}

Here, we present the fitting results for the galaxies U5750, U11454, U11748 and U11819 using both the standard phenomenological DM profiles and the he Pad\'e models.

For the galaxy U5750, the results obtained for the DM standard profiles and Pad\'e approximations are summarized in Table~\ref{tab:U5750}. The best agreement with the observed RC is achieved by Beta, Brownstein and Exponential Sphere profiles ($\Delta\mathrm{BIC}=0$), with equally acceptable choices given by Burkert, ISO, Persic, and Pad\'e~02 ($\Delta\mathrm{BIC}\leq3$), and the remaining Pad\'e~12 and 03 models mildly excluded ($3<\Delta\mathrm{BIC}\leq6$).
All models provide consistent DM halo masses, all in the well determined range of $M=(3.21-3.90)\times10^{10}M_{\odot}$ within the attached errors.

For the galaxy U11454, the best-fit parameters are presented in Table \ref{tab:U11454}. The best-fitting model is the ISO profile ($\Delta\mathrm{BIC}=0$), while the Pad\'e~20 also provides an acceptable description of the observed RC ($\Delta\mathrm{BIC}=2$). The mildly excluded profiles are Burkert, Persic Pad\'e~21 and 30 (($3<\Delta\mathrm{BIC}\leq6$), while the remaining ones are strongly disfavored ($\Delta\mathrm{BIC}>6$). 
The estimated DM halo mass is narrowly determined to stay in the range $M=(5.93-6.76)\times10^{10}M_{\odot}$, with larger uncertainties inferred from the Pad\'e~21 and 30 approximations, however still consistent with the estimate from the best-model ISO.

The results for the galaxy U11748 are shown in Table~\ref{tab:U11748}. 
Contrary to the previous cases, Pad\'e models significantly improve the fits. In particular, the Pad\'e 03 profile yields the best-fit ($\Delta {\rm BIC}=0$), with the Pad\'e~02 and 12 still providing an adequate description of the observed RC ($\Delta {\rm BIC}=3$). All the other models are strongly excluded ($\Delta {\rm BIC}>60$) from a statistical view point.
The DM halo mass estimated is in the range  $M=(1.80-2.76)\times10^{11}M_{\odot}$, where the higher value comes from the best-fit Pad\'e~03 approximation.

Finally, the results for the galaxy U11819 are presented in Table~\ref{tab:U11819}. 
The exponential sphere model provides the preferred fit with $\Delta\mathrm{BIC}=0$. 
All the other phenomenological models are suitable alternatives, being $\Delta\mathrm{BIC}\in[0,1]$.
All the Pad\'e models are mildly excluded, yielding $3<\Delta\mathrm{BIC}\leq6$. 
The estimated DM halo mass corresponding to the best-fit model is $M=6.77\times10^{10}M_{\odot}$.


\subsection{Theoretical discussion}

Overall, the analyses on the adopted data set of galaxies show a preference for phenomenological density profiles, whereas only in the case of U11748 the Pad\'e~03 profile stands out.
Among the phenomenological profiles ISO and exponential sphere are the best-suited models; specifically the ISO is the best fit in the galaxies ESO0140040 and U11454, the exponential sphere is the best model for the galaxy U11819, and in the remaining ones both models are equally viable.
Another important outcome is that the inferred DM masses remain broadly consistent across different parameterizations, including the Pad\'e models.

The main theoretical advantage of the Pad\'e construction is its ability to interpolate between different radial regimes with a small number of parameters. Conventional profiles typically impose a specific asymptotic behavior by construction. For example, the ISO and Burkert profiles are cored, whereas cusped profiles derived from cosmological simulations exhibit a different inner structure. The Pad\'e density laws are not tied to a single predefined asymptotic form. Instead, the ratio of polynomials allows the profile to adjust both the central slope and the outer decline through the fitted coefficients.
Back to our results, the galaxy U11748 represents the clearest example of the potential advantage of the Pad\'e approach.
Here, the Pad\'e~03 model exactly captures all the above features.
The additional coefficients permit a more accurate reconstruction of the radial transition between the inner and outer halo regions only in the specific data set of U11748 that clearly displays the flattening of the RC at large distances, while in all the other cases the RCs stops well before, making difficult to constrain the additional parameters of the Pad\'e approximations (see Figs.~\ref{fig:RC1}-\ref{fig:RC2}).

The above conclusion seems to be confirmed by all the galaxies where the Pad\'e~20 model provides an acceptable fit to the observed RCs.
In all these cases it has been inferred $b_1\approx0$ and $b_2\approx1$, for which the density degenerates with the ISO model $\rho\propto(1+x^2)^{-1}$.
This can be interpreted as the attempt of the Pad\'e models to shape the RCs at larger distances where, however, data lack and, thus, degenerating with the best-fitting phenomenological model ISO.

Additionally, several points should be emphasized. First, the present analysis assumes spherical symmetry and neglects the detailed baryonic decomposition of the galaxies. Second, the Pad\'e profiles introduced here are purely phenomenological. They are not derived from a specific DM particle model, from collisionless N-body simulations, or from a self-consistent solution of the Einstein–Boltzmann system. Third, a broader statistical study involving galaxies with different morphological types, luminosities, and surface brightnesses is required to assess the general applicability of the proposed parameterization. Such an analysis would clarify whether the observed improvement represents a generic property of Pad\'e-based profiles or is primarily associated with the particular characteristics of the galaxies examined in this work.
Despite these limitations, the results demonstrate that the parameterizations of type Pad\'e constitute a useful analytic framework for the modeling of RC.

\section{Final outlooks and perspectives}
\label{conclusion}

In this work, we investigated whether low-order Pad\'e rational functions could be used as \emph{empirical density profiles for DM halos}, through a direct analysis on galaxy RCs. 

The analysis was carried out approximating the halo as fully dominated by DM, namely neglecting the action of gas, baryons, etc., for eight galaxies, namely ESO0140040, ESO3050090, ESO4250180, ESO4880049, U5750, U11454, U11748, and U11819. Three Pad\'e-inspired density profiles, denoted Pad\'e 02, Pad\'e 12, and Pad\'e 03, have been thus introduced and compared with a set of standard phenomenological halo parameterizations, including the ISO, Burkert, Beta, Brownstein, exponential-sphere, and Persic models. The model parameters were inferred through a Metropolis-Hastings MCMC analysis and the relative statistical performance of the different profiles was assessed by means of BIC bounds.

Hence, we tested whether rational density laws could provide a useful phenomenological framework for RC fitting. In this sense, the Pad\'e construction offered a way of introducing additional radial flexibility, through empirical parameters. In this respect, differently from conventional profiles, which typically impose a fixed inner or outer behavior by construction, the rational form allowed the fitted density law to adjust the transition between the inner and outer halo regions through a small number of free coefficients. 

We concluded from our analyses that this feature was particularly useful in assessing whether the observed RCs required more structure than that encoded in the usual two-parameter halo models.

For the ESO subsample, the results showed that the Pad\'e profiles were generally viable, although they were not systematically preferred over the standard phenomenological models. In the case of ESO0140040, the pseudo-isothermal profile provided the best statistical fit among the models considered. Nevertheless, the Pad\'e 02 profile remained statistically acceptable and led to a halo mass compatible with that inferred from the best standard model. The higher-order Pad\'e 12 and Pad\'e 03 profiles did not improve the Bayesian evidence for this galaxy, mainly because their additional freedom was not sufficiently constrained by the available radial information.

For ESO3050090, several standard phenomenological profiles gave statistically comparable descriptions of the observed RC. The Pad\'e 02 profile also provided an acceptable fit, while the Pad\'e 12 and Pad\'e 03 profiles were penalized by the information criterion. This behavior indicated that, for this galaxy, the observed data did not require the additional flexibility associated with higher-order rational profiles. The inferred halo masses remained consistent across the different models, with the exception of cases in which the additional Pad\'e parameters were weakly constrained.

For ESO4250180, a similar pattern emerged. The Pad\'e 02 and Pad\'e 12 profiles were close to the best models in terms of Bayesian evidence, whereas Pad\'e 03 was less favored. The large uncertainties associated with the Pad\'e halo masses showed that the present data did not constrain all rational-profile coefficients with the same efficiency. 

For ESO4880049, the Burkert, exponential sphere, ISO, and Persic parameterizations provided the best statistical descriptions. The Pad\'e 02 profile remained acceptable, but the higher-order Pad\'e 12 and Pad\'e 03 models were again penalized by the BIC. The inferred masses obtained from the Pad\'e fits were compatible in order of magnitude with those of the standard profiles, but the statistical evidence did not require replacing the simpler phenomenological models. 

The U subsample displayed a more differentiated behavior. For U5750, the Beta, Brownstein and exponential-sphere profiles provided the best statistical fits, while the Burkert, ISO, Persic, and Pad\'e 02 profiles remained acceptable alternatives. The higher-order Pad\'e 12 and Pad\'e 03 profiles did not yield a statistically significant improvement. This result showed that, when the RC is already well described by simple cored or exponentially declining profiles, the extra Pad\'e coefficients do not necessarily improve the model selection.

For U11454, the ISO profile gave the best fit. The Pad\'e 02 profile nevertheless remained statistically close to the preferred model, while Pad\'e 12 and Pad\'e 03 were still moderately compatible. This case was particularly useful for understanding the relation between Pad\'e 02 and the ISO profile. Indeed, when the fitted coefficients approach the regime in which the rational denominator effectively reproduces a quadratic behavior, the Pad\'e 02 profile becomes close to a ISO-like form. This degeneracy explains why Pad\'e 02 can provide acceptable fits even when the data do not demand a genuinely new radial structure.

The most relevant result of the analysis was found for U11748 galaxy. There, the standard phenomenological profiles performed significantly worse than the Pad\'e models. The Pad\'e 03 profile gave the lowest BIC and therefore represented the best description among the profiles considered. Pad\'e 02 and Pad\'e 12 also remained suitable, while the conventional models were strongly disfavored. This result indicated that the RC of U11748 contains radial information that cannot be efficiently captured by the simpler two-parameter profiles adopted here. The additional coefficients of the Pad\'e 03 model allowed a better reconstruction of the transition between the inner and outer halo regions. U11748 therefore provided the clearest example in the present sample of the potential advantage of Pad\'e-inspired rational density laws.

For U11819, the exponential sphere model gave the preferred fit, while the other conventional profiles remained statistically close. The Pad\'e profiles were mildly disfavored by the BIC, although they still reproduced the RC at a phenomenological level. This confirmed that the Pad\'e framework is not automatically favored for every galaxy.

Our analysis also showed that the statistical role of the Pad\'e coefficients is strongly data-dependent. When the RC covers only a limited radial range, the extra coefficients are weakly constrained and the BIC naturally penalizes the additional parameter freedom. In these cases the Pad\'e profiles tend either to reproduce the behavior of simpler models or to yield broad posterior uncertainties. Conversely, when the data contain sufficient information on both the rising and flattening portions of the curve, as in U11748, the rational structure becomes more effective. This provides a useful diagnostic criterion for future applications of Pad\'e profiles to larger galaxy samples.

From the limitations of the present work, we intend to perform future developments toward the use of empirical Pad\'e expansion in galaxies. First, the analysis was performed under the assumption of spherical symmetry and did not include an explicit baryonic decomposition. Stellar disks, gas components, possible bulges, and uncertainties in the stellar mass-to-light ratio were not modeled separately. Hence, we intend to work them out in future efforts, testing their impact in our employed profiles. Furthermore, we will study extended catalogs of galaxies and explore to give an explanation to the universal RC existence through the role played by the Pad\'e expansions. 

Last but not least, further developments will also include a systematic comparison with simulations and with non-parametric reconstructions of the halo density. It will be important to quantify how stable the Pad\'e coefficients are under changes of radial range, priors and baryonic assumptions. Finally, the relation between the Pad\'e coefficients and their physical interpretations will be object of future studies for classifying the diversity of observed galactic RCs.

\section*{Acknowledgments}
This research was funded by the Science Committee of the Ministry of Science and Higher Education of the Republic of Kazakhstan (Grant No. AP23488743).

\bibliographystyle{unsrt}
\bibliography{0rrr}
\end{document}